\documentclass[12pt]{article}
\usepackage[margin=1 in]{geometry}
\usepackage{natbib}
\usepackage{float}
\usepackage{cite}
\usepackage[colorlinks=true,citecolor=blue,breaklinks]{hyperref}
\usepackage{graphicx}
\usepackage{amssymb}
\usepackage{url}
\usepackage{subcaption}
\usepackage{mathrsfs}
\usepackage{amsmath}

\begin{document}

\title{Influence of flow thickness on general relativistic low angular momentum accretion around spinning black holes}
\author{Pratik Tarafdar$^1$\footnote{pratikta16@gmail.com}, Susovan Maity$^2$\footnote{susovanmaity@hri.res.in}, Tapas K. Das$^2$\footnote{tapas@hri.res.in} \\
\\
$^1$S. N. Bose National Centre for Basic Sciences, \\
Block JD, Sector III, Salt Lake, Kolkata 700 106, India. \\
$^2$Harish-Chandra Research Institute, HBNI, \\ 
Chhatnag Road, Jhunsi, Allahabad 211 109, India. \\} 
\date{}

\maketitle

\begin{abstract}
\noindent
General relativistic, axisymmetric flow of low angular momentum accretion around a Kerr black hole can have certain geometric configuration where the flow is maintained in
hydrostatic equilibrium along the vertical direction (direction orthogonal to the equatorial plane of the flow). The flow thickness for such accretion models becomes a function
of the local radial distance measured from the black hole horizon. There are three types of functions defined in the literature which resemble the thickness of the flow for such
a configuration. We formulate the equations governing the steady state astrophysical accretion characterized by both the polytropic as well as the isothermal
equation of state in classical thermodynamics. We solve the equations within the framework of such geometric configuration for three different thickness functions, to obtain the multi-transonic,
shocked, stationary integral accretion solutions. Such solutions enable us to study how flow thickness influences the dependence of the properties of post-shock flows on black hole spin angular
momentum, i.e., the Kerr parameter. For temperature-preserving standings shocks, we find that the post-shock part of the disc can become luminous and considerable amount
of gravitational energy carried by the accreting fluid can get liberated at the shock. We find which kind of thickness function produces the maximum liberated energy, making the disc most luminous.
\end{abstract}

\section{Introduction}

\noindent
Axially symmetric, low angular momentum accretion of hydrodynamic fluid onto astrophysical
black holes may exhibit multi-transonic features, and such multi-transonic accretion flow
is endowed with a stationary shock. Such low angular momentum, practically inviscid flow may be observed in realistic astrophysical systems like detached binaries fed by accretion from OB stellar winds (\citep{is75aa}, \citep{ln84ssr}), semi-detached low-mass non-magnetic binary systems (\citep{bbck98mnras}), supermassive black holes fed by accretion  from weakly rotating central stellar clusters (\citep{illarionov88sa}, \citep{ho99} and references therein). For a standard Keplerian accretion flow, various physical processes like turbulence, produce practically inviscid low angular momentum flow (see, e.g. \citep{ia99mnras} and references therein). Several recent works on accretion onto our Galactic Centre black hole indicates the presence of such flow as well (\citep{melia92apj}, \citep{mlc01apj}, \citep{mdc06mnras}, \citep{moscibrodzka06aa}, \citep{cmpds07RAGtime}, \citep{mmzr07apj}, \citep{gswlddmmybkmn08apj}, \citep{gefbdpog09apj}, \citep{ferrier09aap}, \citep{geg10revmp}, \citep{om12mnras}). \\

\noindent
The multi-transonic features and the formation of the corresponding standing shock have
been studied extensively by several authors in the last forty years. Such efforts were initiated
for black hole accretion under the influence of the post-Newtonian pseudo-Schwarzschild
and pseudo-Kerr potential (\citep{az81apj}, \citep{mp82aa}, \citep{fukue83pasj}, \citep{lu85aa}, \citep{mc86aa}, \citep{blaes87mnras}, \citep{chakrabarti89ApJ}, \citep{nf89pasj}, \citep{ak89apj}, \citep{ac90apj}, \citep{sm94mnras}, \citep{btb97prl}, \citep{tkb98aa}, \citep{cd01mnras}, \citep{das02apj}, \citep{dpm03apj}, \citep{ottm04pasj}, \citep{fukue04pasja}, \citep{fukue04pasj}, \citep{mdc06mnras}, \citep{dc12mnras}, \citep{ssnrd16na}, \citep{mnd18mnras}, \citep{ddmc18prd}). Eventually, shocked multi-transonic flows have
been studied for general relativistic accretion flows as well (\citep{lu86grg}, \citep{ky94mnras}, \citep{yk95aa}, \citep{pariev96mnras}, \citep{nayakama96mnras}, \citep{chakrabarti96mnras}, \citep{chakrabarti96apj}, \citep{pa97mnras}, \citep{ft04apj}, \citep{bdw04apjl}, \citep{ny08apj}, \citep{ny09apj}, \citep{dc12na}, \citep{dnhbmcbwkn15na}, \citep{sj15mnras}, \citep{scj17mnras}, \citep{sukova17RAGtime}, \citep{td18na}, \citep{tbnd19prd}, \citep{pjs19mnras}, \citep{ddmn19mnras}, \citep{ddn19mnras}, \citep{fukue19mnras}), where the work
by Fukue (\citep{fukue83pasj}, \citep{fukue87pasj}) may be attributed to the first ever paper published in the field
of study of multi--transonic shocked accretion flow. Of late, such shocked flows have been
studied for magneto--hydrodynamic black hole accretions as well (\citep{trft92apj}, \citep{trft02apj}, \citep{takahashi02apj}, \citep{tgfrt06apj}, \citep{ftt07apj}, \citep{sd18jaa}). \\

\noindent
Geometrical configuration of axisymmetrically accreting fluid can assume three different forms, see e.g., section $4$ of \citep{bcdn14cqg}, for detailed discussions on
such configurations. Also see \citep{cd01mnras}, \citep{tbnd19prd} and references therein. In the present work, we concentrate
on axially symmetric flows under hydrostatic equilibrium in the vertical direction, where the
gravitational force on the accreting fluid is balanced against the fluid pressure force.
These commonly used disc models, however, possess certain limitations, and there are certain
proposals available in the literature to calculate a more realistic expression for the disc
thickness, e.g., \citep{beskin97pu}, \citep{hh98apj}, \citep{bt05aa}, \citep{dh06apjs} and \citep{beskin09}. We, nevertheless, stick to the disc structure maintained in hydrostatic
equilibrium along vertical directions, since dealing with the aforementioned alternate disc models
is mathematically very difficult, if not impossible, while obtaining the stationary integral flow solutions from
general relativistic Euler and the continuity equations. \\

\noindent
First ever detailed calculation of flow structure for general relativistic accretion onto rotating
black holes was obtained by Novikov and Thorne (hereafter NT). They provided a particular
expression of the disc thickness for flow in hydrostatic equilibrium along the vertical
direction (\citep{nt73}). \\

\noindent
Such an expression was slightly modified by Riffert and Herold (\citep{rh95apj}, hereafter RH) because the later work directly used the general relativistic Euler equation to derive the gravity-pressure balance equation, whereas in the first work the general relativistic version of the gravity-pressure balance equation was not directly derived. NT took the Newtonian gravity-pressure balance equation and replaced the vertical component of gravity pressure balance with $R^z_{0z0} z$. \\

\noindent
In recent years, Abramowicz, Lanza and Percival (\citep{alp97apj}, hereafter ALP) have provided
a novel expression for disc thickness. In these calculations, ALP also derived the same gravity-pressure balance equation from the general relativistic equation. But while simplifying the equation they replaced the four velocity component in such a way that no singularity in the disc height occurs at horizon. The main modification apart from this careful choice of four velocity is that ALP used only one component of the relativistic Euler-equation whereas RH did not assume trivial forms of four velocities and solved two equations simultaneously for two components of the relativistic Euler equation. \\

\noindent
In our present work, we will formulate and solve the general relativistic Euler and the continuity
equations to observe how the aforementioned three different prescriptions for the flow
thickness influence the properties of the stationary integral flow solutions having more than
one sonic transitions and incorporating standing shock. Accretion flow governed by the polytropic as well as the isothermal equation of state will be studied. \\

\noindent
We shall learn that for shock formation in isothermal flow, considerable
amount of energy may be released at the shock, which may enhance the brightness of the otherwise advection-dominated
radiatively inefficient disc near the shock and such mechanism may explain the
details of the flares emanating out of the black hole accretion disc as observed in various
wavebands of the electromagnetic spectrum (\citep{kbevdkh17mnras}, \citep{mg17aa}, \citep{rjwo17mnras}, \citep{mfbv20aa}). \\

\noindent
Our work, thus, sheds light on
how a proposed flow thickness may contribute to understand the variation of the disc luminosity
during the generation of flares. The present work may also be useful in the context of
analogue gravity phenomenon. It has been observed that a curved acoustic metric may be
embedded within the accreting matter and such space-time may be generated through the
perturbation of accretion flow (\citep{abd15grg}, \citep{abd17na}, \citep{sfd17cqg}, \citep{sd18prd}, \citep{shaikh18cqg}, \citep{dsd18na}, \citep{smnd19na}, \citep{dd19arxiv}). The present work will also lead to the understanding
of how the flow thickness of axially symmetric accretion in the Kerr metric may
influence the properties of the analogue surface gravity of the corresponding sonic space-time. \\

\noindent
Overall, the technical procedures followed to accomplish our goal are summarized below -- \\

\noindent
For three expressions of the flow thickness as classified in previous paragraphs, we shall
formulate and solve the general relativistic Euler equation and the equation of continuity for
ideal relativistic fluid, by assuming that the viscous transport of angular momentum may
not play significant role for low angular momentum accretion flow. We shall then solve such
equations for steady state flows and obtain stationary integral flow solutions which may
make transitions from subsonic to supersonic state twice. We then introduce and discuss the
mathematical conditions governing the formation of general relativistic standing shock, and
solve such equations to obtain the shock location as a function of black hole spin angular
momentum, i.e. the Kerr parameter. The properties of hotter, denser and shock-compressed
post-shock flow is then studied as the function of the Kerr parameter and the influence of
the expression to the flow thickness on such properties is then realized.

\section{Model}

\subsection{Physical space-time}

\noindent
We represent the physical space-time of an uncharged, rotating black hole along its equatorial
plane using the Kerr metric written in cylindrical Boyer-Lindquist coordinates (\citep{bl67jmp}). 
The choice of co-ordinates is in accordance with cylindrical symmetry of the discs. Also for simplicity,
we are interested in projection of the flow variables on the equatorial plane, obtained using
vertical averaging technique as explained in subsequent sections. The line element for such
a metric is given by,

\begin{equation}
ds^2=-\frac{r^2\Delta}{A}dt^2+\frac{A}{r^2}(d\phi-\omega dt)^2+\frac{r^2}{\Delta}dr^2+dz^2,
\label{background_metric}
\end{equation}

\noindent
where 
\begin{equation}
\Delta=r^2-2r+a^2, \quad A=r^2+r^2a^2+2ra^2, \quad \omega=\frac{2ar}{A}.
\end{equation}

\noindent
$\omega$ represents the rate of frame dragging by the black hole, $a$ being the Kerr parameter which
in turn is related to the spin angular momentum $J$ of the black hole through the relation 
$a=J/M_{BH}c$, where $-1 < a < 1$, $M_{BH}$ is the mass of
the respective black hole and $c$ is the velocity of light in vacuum. Calculations have been carried out using natural units, i.e.
$G=c=1$ where $G$ is the universal gravitational constant. All masses are measured in units of $M_{BH}$ which has been set to $1$ for algebraic
convenience and can be easily substituted back using simple dimensional analysis. Distances
are measured in units of $GM_{BH}/c^2$, times are measured in units of $GM_{BH}/c^3$ and all velocities
are scaled in units of $c$. For a Kerr black hole, the horizon is
located at the outer boundary of $g^{rr}= \Delta/r^2 = 0$, which is defined as $r_+$ and the expression of which is given by,

\begin{equation}
r_+ = 1+\sqrt{1-a^2}.
\label{r_plus}
\end{equation}  

\subsection{Choice of disc height}

\noindent
We consider accretion disc around Kerr black hole in hydrostatic equilibrium along vertical
direction, i.e, the gravitational force component is balanced by the pressure of the fluid
constituting the disc. The earliest general relativistic formulation of this gravity-pressure
balance and thus a vertically averaged height prescription proposed by NT
(\citep{nt73}) is given by,

\begin{equation}
H_{NT} (r)= \left(\frac{p}{\rho}\right)^\frac{1}{2} \frac{r^3 + a^2 r + 2a^2 }{r^{\frac{3}{2}} + a} \sqrt{\frac{r^6 - 3r^5 + 2ar^{\frac{9}{2}}}{(r^2 -2r +a^2)(r^4 + 4a^2 r^2 -4a^2 r + 3a^4 )}},
\label{NTheight}
\end{equation}

\noindent
where $p$ and $\rho$ are pressure and rest-mass energy density of the fluid respectively. It is to be noted that accretion flow described by the above disc thickness can not be extended upto $r_+$. The flow will be truncated at a \textit{truncation radius} $r_T$, which is given by solution of the equation,

\begin{equation}
\left( r_T \right)^{\frac{1}{2}} (r_T -3) = 2a,
\label{r_truncated}
\end{equation}

\noindent
and which is greater than $r_+$. In reality of course the flow will exist upto $r_+$, but stationary integral flow solutions can not be formulated in the vicinity of $r_+$ for NT-type of discs. \\

\noindent
The next prescription found in literature dealing with gravity-pressure balance and proposing a height recipe in the Kerr metric was by RH (\citep{rh95apj}). They modified the gravity-pressure balance condition of the treatment done in NT. Their proposed disc height is given by,

\begin{equation}
H_{RH} (r) = \left(\frac{p}{\rho}\right)^\frac{1}{2} \sqrt{\frac{r^5 - 3r^4 + 2ar^{\frac{7}{2}}}{r^2 -4ar^{\frac{1}{2}} + 3ar^2}}
\label{RHhieght}
\end{equation}

\noindent
Here also, the flow can only be extended inwardly upto $r_T$, which has the same value for NT and RH discs around the same black hole as given by eqn.(\ref{r_truncated}). \\

\noindent
Thus we see that both the
disc heights can be expressed in the form by $H(r)=\left(\frac{p}{\rho}\right)^\frac{1}{2}f(r,a)$. The difference between these two
models of disc thickness in vertical equilibrium is reflected by the difference in functional
form of two different $f(r,a)$. The essential difference arises because whereas NT
balanced the vertical component of pressure with a particular Riemann tensor
$R^z_{0z0}$, which was equivalent to the vertical component of gravitational acceleration, RH derived the gravity-pressure balance equation by simultaneously solving two orthogonal projection components of the general relativistic Euler equation. We will
observe that the Mach number evaluated at the critical points corresponding to the flow described
by the thickness function proposed by NT will be identical with that
of the flow described by the thickness functions proposed by RH. This is evident because
the dynamical equation will have the same form in terms of $f(r,a)$, because the height recipe also has a similar form. \\

\noindent
ALP (\citep{alp97apj}) introduced an expression for the disc thickness, given by

\begin{equation}
H_{ALP} (r) = \left(\frac{p}{\rho}\right)^\frac{1}{2} \sqrt{\frac{2r^4}{v_\phi^2 - a^2 (v_t - 1)}}.
\label{ALPheight}
\end{equation}

\noindent
Here, $v^\mu$ denotes the four-velocity of the fluid in an azimuthally-boosted frame that co-rotates with the flow.
$v_{\phi}$ and $v_t$ are the azimuthal and temporal components of the covariant 4-velocity respectively which are related by 
$\lambda=-v_{\phi}/v_t$, where $\lambda$ is the specific angular momentum of the flow and $v_t$ is given by,

\begin{equation}
v_{t} = \sqrt{\frac{\Delta}{B(1-u^2)}},
\label{v_t}
\end{equation}

\noindent
where $ B = g_{\phi\phi} + 2\lambda g_{t\phi} -\lambda^2 g_{tt} $ and $u$ denotes advective velocity which is the three-component velocity in the co-rotating frame.\footnote{we refer \citep{gp98apj} for the detailed description of expressions of various velocities in different frames
for rotating accretion flow in the Kerr metric}
As mentioned earlier, no singularities in ALP-type disc heights occur at the horizon. Thus, ALP discs do not have any truncation constraints and the steady state accretion solutions can be obtained upto $r_+$.

\section{Polytropic accretion}

\subsection{Fluid equations}

\subsubsection{Fluid specification and sound speed}

\noindent
As specified earlier, we consider a low angular momentum accretion disc. The low angular momentum prevents the inward part of the disc to transfer momentum to the outside region. Thus we cosider a perfect fluid as the constituent of the accretion disc. The energy momentum tensor for a perfect fluid is given by

\begin{equation}
T^{\mu\nu} = (p+\epsilon)v^\mu v^\nu + p g^{\mu\nu} ,
\label{energy-mom-tensor}
\end{equation}

\noindent
where $\epsilon$ is the total energy density of the fluid given by $ \epsilon = \rho + \epsilon_{\rm thermal} $, where $\epsilon_{\rm thermal}$ is the internal thermal energy density of the fluid. \\

\noindent
The equation of state for adiabatic flow is given by $ p = k\rho^\gamma $ where $\gamma$ is the polytropic index and $ k $ is a constant. Whereas for isothermal case $ p\propto \rho $. The sound speed for adiabatic flow (isoentropic flow) is given by

\begin{equation}
c_{s}^2 = \left.\frac{\partial p}{\partial \epsilon}\right|_{\rm entropy} = \frac{\rho}{h}\frac{\partial h}{\partial \rho} ,
\label{cs-ad}
\end{equation}

\noindent
where $ h $ is the enthalpy given by

\begin{equation}
h = \frac{p+\varepsilon}{\rho}.
\label{enthalpy}
\end{equation}

\subsubsection{Conservation of specific energy}

\noindent
The energy-momentum conservation equation can be written as

\begin{equation}
D_\mu T^{\mu\nu} = 0,
\label{energy-mom-con}
\end{equation}

\noindent
where $D_\mu$ is the covariant derivative operator with respect to $\mu$. Eqn.(\ref{energy-mom-con}), in turn, can be written using the definition of sound speed as
\begin{equation}
(p + \epsilon) v^\mu D_\mu v^\nu +(v^\mu v^\nu + g^{\mu\nu})\partial_\mu p = 0.
\label{Euler}
\end{equation}

\noindent
Now the thermodynamic equation of motion is given by
\begin{equation}
T \partial_\mu s = \partial_\mu h -\frac{\partial_\mu p}{\rho}
\label{thermodynamic_eq}
\end{equation}

\noindent
where $s$ is the specific entropy. In case of polytropic accretion, right hand hand side of Eqtn. (\ref{thermodynamic_eq}) is zero and Eq. (\ref{Euler}) can be rewritten using normalization of four velocity, which yields
\begin{equation}
u^\nu \left[D_\nu( hu_\mu) - D_\mu(hu_\nu) \right] = 0
\end{equation}

\noindent
Using time component of the equation and the fact that the flow is stationary, the conserved quantity from energy-momentum conservation equation in case of polytropic accretion turns out to be
\begin{equation}
\mathcal{E} = hv_{t} = {\rm constant}.
\end{equation}

Substituting for $v_t$ from eqn.(\ref{v_t}) and $h$ from eqn.(\ref{enthalpy}) we obtain,

\begin{equation}
\mathcal{E} = \frac{\gamma -1}{\gamma -1-c_{s}^2}\sqrt{\frac{\Delta}{B(1-u^2)}}
\label{xi}
\end{equation}

\noindent
Taking logarithmic derivative of both sides of equation (\ref{xi}) gives the gradient of sound speed as
\begin{equation}
\frac{dc_{s}}{dr} = -\frac{\gamma -1-c_{s}^2}{2c_{s}}\left[\frac{u}{1-u^2}\frac{du}{dr}+\frac{1}{2}\left(\frac{\Delta'}{\Delta}-\frac{B'}{B}\right)\right]
\label{dcdr}
\end{equation}

\subsubsection{Conservation of mass}

\noindent
The mass conservation equation is given by

\begin{equation}\label{continuity}
D_\mu (\rho v^\mu) = 0.
\end{equation}
  
\noindent
A vertical averaging is done for convenience by integrating the flow equations over the $z$ co-ordinate and the resultant equation is described by the flow variables defined on the equatorial plane ($ z = 0 $). Furthermore, integration is done over $ \phi $ which gives a factor of $ 2\pi $ due to the axial symmetry of the flow. We apply such vertical averaging as prescribed in (\citep{nt73}, \citep{mkfo84pasj}, \citep{gp98apj}) to the continuity equation given by Eq. (\ref{continuity}). The vertically averaged $z$-component of the 4-velocity becomes $ v^z \sim 0 $. Thus for the stationary ($ t $-independent) and axially symmetric ($ \phi $-independent) flow, the continuity equation turns out to be

\begin{equation}\label{continuity-avg-st}
\frac{\partial}{\partial r}(4\pi H_\theta\sqrt{-g}\rho v^r) = 0
\end{equation}

\noindent
$ H_\theta $ arises due to the vertical averaging and is the local angular scale of flow. One can relate the actual local flow thickness $ H(r) $ to the angular scale of the flow $ H_\theta $ as $ H_\theta = H(r)/r $, where $ r $ is the radial distance along the equatorial plane from the centre of the disc. $g$ is the value of the determinant of the metric $ g_{\mu\nu} $ on the equatorial plane, $g = {\rm det}(g_{\mu\nu})|_{z=0} = -r^4$. The equation (\ref{continuity-avg-st}) gives the mass accretion rate $ \dot{M} $ as

\begin{equation}\label{Sationary-mass-acc-rate}
\dot{M} = 4\pi \sqrt{-g}H_\theta  \rho v^r = 4\pi H(r) r \rho v^r.
\end{equation}

\noindent
The $ r $ component of the four velocity, $ v^r $ is related to $ u $ by the transformation law as

\begin{equation}\label{transformation}
v^r = \frac{u}{\sqrt{g_{rr}(1-u^2)}} = \frac{\sqrt{\Delta}u}{r\sqrt{1-u^2}}
\end{equation}

\noindent
using $ g_{rr}=r^2/\Delta $, $ \dot{M} $ can be written as
\begin{equation}\label{Psi}
\dot{M} = 4\pi H(r)\Delta^{1/2}\rho \frac{u}{\sqrt{1-u^2}} = {\rm constant}
\end{equation}

\noindent
For adiabatic flow, a new quantity $ \dot{\Xi }$ is obtained from $ \dot{M} $ by multiplying it with $ (\gamma k)^{\frac{1}{\gamma-1}} $. $ \dot{\Xi} $ is a measure of entropy accretion rate and typically called as the entropy accretion rate. The concept of the entropy accretion rate is widely used in accretion astrophysics. The entropy accretion rate was first defined in the literature by \citep{az81apj} and \citep{blaes87mnras}. Expressing $ \rho $ in terms of $ \gamma, k$ and $c_{s} $ gives

\begin{equation}\label{Xi}
\dot{\Xi} = \left(\frac{(\gamma -1)c_{s}^2}{\gamma -1- c_{s}^2}\right)^{\frac{1}{\gamma -1}} 4\pi H(r)\Delta^{1/2} \frac{u}{\sqrt{1-u^2}} = {\rm constant}
\end{equation}

\noindent
To express the entropy density in terms of $u$ , $c_{s}$ and $r$ only, the expression of height must be written in terms of $u$ and $c_{s}$ also. For this we first note that, for adiabatic equation of state, $ p/\rho $ can be written as
\begin{equation}
\frac{p}{\rho} = \left(\frac{1}{\gamma }\right)\left(\frac{(\gamma -1)c_{s}^2}{\gamma -1-c_{s}^2}\right)
\end{equation}

\noindent
This factor is common to all the height recipes. Now, for convenience, we distinguish the height recipes in two classes, one consisting of NT and RH. The other has ALP model as its member. The reason behind this classification is that whereas the models in the first class can be written in generally as $H(r)= (\frac{p}{\rho})^{\frac{1}{2}}f(r,a)$ , the model in the other class can not be written as such. Thus we proceed seperately for this two classes and derive the desired velocity gradients. \\

\noindent
\paragraph{NT and RH-type of discs}
For these two models, we can write $ H(r) $ from Eq. (\ref{NTheight}) and Eq.(\ref{RHhieght}) as
\begin{equation}\label{NTheightshort}
H(r) = \left(\frac{1}{\gamma }\right)^{1/2}\left(\frac{(\gamma -1)c_{s}^2}{\gamma -1-c_{s}^2}\right)^{1/2}f(r,a)
\end{equation}

\noindent
where for NT
\begin{equation}
f_{NT}(r,a) = \frac{r^3 + a^2 r + 2a^2 }{r^{\frac{3}{2}} + a} \sqrt{\frac{r^6 - 3r^5 + 2ar^{\frac{9}{2}}}{(r^2 -2r +a^2)(r^4 + 4a^2 r^2 -4a^2 r + 3a^4 )}}.
\label{f_NT}
\end{equation}

\noindent
and for RH
\begin{equation}\label{f_RH}
f_{RH} (r) =  \sqrt{\frac{r^5 - 3r^4 + 2ar^{\frac{7}{2}}}{r^2 -4ar^{\frac{1}{2}} + 3ar^2}}
\end{equation}

\noindent
Using the expression of $ H(r) $ for both these models, $ \dot{\Xi} $ can be written as
\begin{equation}
\dot{\Xi} = \sqrt{\frac{1}{\gamma }}\left(\frac{(\gamma -1)c_{s}^2}{\gamma -1-c_{s}^2}\right)^\frac{\gamma +1}{2(\gamma -1)}4\pi\Delta^{1/2} \frac{u}{\sqrt{1-u^2}}f(r,a)
\end{equation}

\noindent
Taking logarithmic derivative of both sides of the above equation and substituting $ dc_{s}/dr $ using Eq. (\ref{dcdr}) gives
\begin{equation}\label{dudr}
\frac{du}{dr} = \frac{u(1-u^2)\left[\frac{2}{\gamma +1}c_{s}^2(\frac{\Delta'}{2\Delta}+\frac{f'}{f})+\frac{1}{2}(\frac{B'}{B}-\frac{\Delta'}{\Delta})\right]}{u^2-\frac{c_s^2}{(\frac{\gamma +1}{2})}}=\frac{N}{D}.
\end{equation}

\noindent
\paragraph{ALP-type of discs}
From eqn. (\ref{ALPheight}) and using the relation $\lambda = -\frac{v_{\phi}}{v_t}$, we have

\begin{equation}\label{ALPheightmodified}
H(r) = H_{ALP} (r) = \left(\frac{1}{\gamma }\right)^{1/2}\left(\frac{(\gamma -1)c_{s0}^2}{\gamma -1-c_{s0}^2}\right)^{1/2} \sqrt{\frac{2r^4}{\lambda^2 v_t^2 - a^2 (v_t - 1)}} ,
\end{equation}

\noindent
Thus using this expression of $H(r)$, $\dot{\Xi}$ can be expressed as
\begin{equation}
\dot{\Xi} = \sqrt{\frac{1}{\gamma }}\left(\frac{(\gamma -1)c_{s}^2}{\gamma -1-c_{s}^2}\right)^\frac{\gamma +1}{2(\gamma -1)}4\pi\Delta^{1/2} \frac{u}{\sqrt{1-u^2}}\sqrt{\frac{2r^4}{\lambda^2 v_t^2 - a^2 (v_t - 1)}}
\end{equation}

\noindent
Taking logarithmic derivative of the entropy accretion rate and using Eq. (\ref{dcdr}) to replace $\frac{dc_s}{dr}$ we yield
\begin{equation}
\frac{du}{dr}=\frac{\frac{2 c_s^2}{\gamma +1}\left(-\frac{Pv_t \left(2 \lambda ^2 v_t-a^2\right)}{4 F}+\frac{\Delta '}{2 \Delta }+\frac{2}{r}\right)-\frac{P}{2}}{\frac{u}{1-u^2}-\frac{2 c_s^2}{\gamma +1}\frac{1}{\left(1-u^2\right)u}{\left(1-\frac{u^2 v_t \left(2 \lambda ^2 v_t-a^2\right)}{2 F}\right)}} = \frac{N}{D}
\end{equation}

\noindent
where $P=\frac{\Delta'}{\Delta } - \frac{B'}{B}$ and $F=\lambda^2 v_t^2-a^2(v_t-1)$.

\subsection{Critical point conditions}

\noindent
In this section, we will present the scheme and calculations for finding stationary transonic
flow solutions for all the three height recipes. We present NT and RH discs 
in the first class and ALP discs in the second class for reasons stated earlier.

\subsubsection{NT and RH-type of discs}

\noindent
Borrowing a standard recipe from the theory of dynamical systems (\citep{js99}, \citep{strogatz01},\citep{hilborn01}), we set the numerator
and denominator of $du/dr$ to zero separately in order to obtain the necessary conditions to be
satisfied at the critical points of the system. Setting $D=0$ we get,

\begin{equation}\label{critpointsNTRH}
u^2|_c =c_{s}^2 |_c/(\frac{\gamma +1}{2})
\end{equation}

\noindent
where the suffix $c$ denotes its value evaluated at the critical point. By setting $N=0$, we yield
 
\begin{equation}\label{c_s_critical}
c_{s}^2|_c = \left( \frac{\gamma +1}{4} \right)\frac{\frac{B'}{B}-\frac{\Delta'}{\Delta}}{(\frac{\Delta'}{2\Delta}+\frac{f'}{f})}.
\end{equation}

\noindent
In order to solve for the critical points, the critical point condition (\ref{critpointsNTRH}) is used in (\ref{xi}), which gives

\begin{equation}\label{rc}
\mathcal{E} = \frac{\gamma - 1}{\gamma - 1- (c_{s}^2)_c }\sqrt{\frac{(\gamma + 1)\Delta_c}{B_c\left(\gamma + 1 - 2(c_{s}^2)_c \right)}}
\end{equation}

\noindent
where $(c_{s}^2)_c$ is a function of $r$ and Kerr parameter $a$ obtained from (\ref{c_s_critical}). 
The solutions of this equation for a given set of system parameters 
$\left[\mathcal{E},\lambda,\gamma,a\right]$, provide the critical points.
The number of such critical points may be more than one depending on the parameter values. \\

\noindent
Now the value of $\frac{du}{dr}$ at critical point is obtained by using L'Hospital's rule in (\ref{dudr}) as both the numerator and denominator tends to zero at critical point. Then one obtains a quadratic equation of the form
  \begin{equation}\label{twoslopes}
  \alpha_1 \left( \frac{du}{dr} \right)^2 - \alpha_2 \left( \frac{du}{dr} \right) - \alpha_3 = 0,
  \end{equation}

\noindent
  where
  \begin{equation}
  \alpha_1 = 2(u)_c \left[ 1-\frac{\left( (c_{s}^2)_c -\gamma -1 \right)}{(\gamma +1)(1-(u^2)_c)} \right],
  \end{equation}
 
  \begin{multline}
  \alpha_2 = \frac{\left( (c_{s}^2)_c -\gamma -1 \right)}{(\gamma +1)} \left( \frac{\Delta'}{\Delta} - \frac{B'}{B} + \left( \frac{\Delta'}{\Delta} + \frac{2f_{\rm NT}'}{f_{\rm NT}} \right)(u^2)_c \right)\\  + \left[\frac{2}{\gamma +1}(c^2)_c\left( \frac{\Delta'}{2\Delta}+\frac{f_{\rm NT}'}{f_{\rm NT}}\right) +\frac{1}{2}\left( \frac{B'}{B}-\frac{\Delta'}{\Delta}\right) \right] \left( 1-3(u^2)_c \right),
  \end{multline}
 
  \begin{multline}
  \alpha_3 = (u)_c\left(1-(u^2)_c\right) \left[ \frac{1}{2}\frac{\alpha'}{\alpha}^2 - \frac{\alpha''}{2\alpha} + \frac{2(c_{s}^2)_c}{\gamma + 1}\left( \frac{\Delta''}{2\Delta} + \frac{f^{''}}{f} -\frac{1}{2}\frac{\Delta'}{\Delta}^2 -\frac{f'}{f}^2 \right) \right.\\
  \left. + \frac{\left( (c_{s}^2)_c -\gamma -1 \right)}{(\gamma +1)}\left( \frac{\Delta'}{2\Delta} + \frac{f^{'}}{f} \right) \left( \frac{\Delta'}{\Delta} - \frac{B'}{B} \right) \right].
  \end{multline}

\noindent
The two roots of above quadratic equation (\ref{twoslopes}) signify two different slopes of two different integral solutions passing through each critical point, while such slopes are measured at the respective critical points only. \\

\noindent
Once we are equipped with the critical points and the critical gradients, the phase-space
portrait (i.e., $u$ vs. $r$ diagram) can be plotted by numerically integrating the expression
of $du/dr$ (eqn. (\ref{dudr})) for a particular set of $\left[\mathcal{E},\lambda,\gamma,a\right]$ 
as will be illustrated in the subsequent
sections. Here we substitute the value of $c_s^2$ from eqn. (\ref{Xi}) as a function of parameter $\mathcal{E},$ $r$
and $u$. While addressing transonicity-related aspects of the flow, it is usually convenient to
use the Mach number ($M = u/c_s$) instead of the advective flow velocity $u$.

\subsubsection{ALP-type of discs}

\noindent
By setting $N=0$ and $D=0$, the critical conditions turn out to be

\begin{equation}\label{critALP1}
u^2|_{r_c}=\frac{\text{P}}{\frac{\Delta '}{\Delta }+\frac{4}{r}}|_{r_c},
\end{equation}
and
\begin{equation}\label{critALP2}
c_s^2|_{r_c}=\frac{(\gamma +1) \left(2 F u^2\right)}{2 \left(2 F-u^2 v_t \left(2 \lambda ^2 v_t-a^2\right)\right)}|_{r_c}
\end{equation}

\noindent
To find the critical point we use the critical condition (\ref{critALP2}) in eqn. (\ref{xi}) and then solve the equation. \\

\noindent
As observed from eqns. (\ref{critpointsNTRH}), (\ref{critALP1}) and (\ref{critALP2}), the value of the Mach number at the critical point differs from unity for all three disc models considered in the present work.
Hence the critical points do not coincide with the sonic points -- by definition the sonic point is the location where the Mach number becomes unity. The value of
the Mach number at critical points are found out to be less than unity for all three type of disc thicknesses. For NT and RH kind of flow, the Mach number at the
critical point is a constant $\left(\sqrt{\frac{2}{\gamma+1}}\right)$ for a fixed value of $\gamma$. The departure of the value of the Mach number from unity has the same
numerical value for all three critical points and hence three critical points lie on the same horizontal line parallel to the abscissa.  For isothermal accretion,
the value of the polytropic index $\gamma$ will be one, and hence for isothermal flow, the critical and the sonic points will be the same for NT as well as for
RH-type of flow. \\

\noindent
For ALP-type of disc, however, the amount of departure of Mach number (measured at the critical point) from unity is not constant. It rather depends on the
value of the critical point itself.  Thus for ALP type of disc, three different critical points for multi-critical accretion will assume three different values of  the Mach
number, and three critical points will not lie of the same horizontal line for such disc model. For ALP-type of disc, even for the isothermal flow, the Mach number
does not become unity at the critical point. \\

\noindent
Given a set of values of $\left[{\cal E}, \lambda, \gamma, a\right]$, one obtains the location of the critical point through the critical point analysis and it is not required
to integrate the fluid equations (the Euler equation or the equation of continuity). Among three critical points, the middle one is of centre-type and hence no
physical accretion solution can pass through it. Accretion solution can pass through the inner and outer critical points only. Hence, one can have the sonic points
corresponding to these two critical points since both the inner and the outermost critical points are of saddle-type. One thus computes the location of the critical
point algebraically as discussed in sections $3.2.1$ and $3.2.2$, and then integrates the flow equations, starting from the critical point, up to the value of $r$ where the value of the
Mach number becomes unity. That point is defined as the sonic point. We thus need to construct the integral accretion solution to find out the location of the sonic
points corresponding to the inner-type and the outer-type critical points. \\

\noindent
By following the same procedure as used to derive the slopes of trajectories through critical points, we find
\begin{equation}\label{slopeALP}
\frac{du}{dr}|_{r_c}=-\frac{\beta_{VE}}{2\alpha_{VE}} \pm \frac{1}{2\alpha_{VE}}\sqrt{\beta_{VE}^2-4\alpha_{VE}\Gamma_{VE}}
\end{equation}

\noindent
where the co-efficients $\alpha_{VE}$, $\beta_{VE}$ and $\Gamma_{VE}$ are given by, \\
$\alpha_{VE}=\frac{1+u^2}{\left(1-u^2\right)^2}-\frac{2nD_2D_6}{2n+1}$, $\beta_{VE} =\frac{2nD_2D_7}{2n+1}+\tau_4$, $\Gamma_{VE}=-\tau_3$, \\
$n=\frac{1}{\gamma -1}$, $D_2=\frac{c_s^2}{u\left(1-u^2\right)}\left(1-D_3\right)$, $D_6=\frac{3u^2-1}{u\left(1-u^2\right)}-\frac{D_5}{1-D_3}-\frac{\left(1-nc_s^2\right)u}{nc_s^2\left(1-u^2\right)}$, \\
$D_7=\frac{1-nc_s^2}{nc_s^2}\frac{P1}{2}+\frac{D_3D_4v_tP1}{2\left(1-D_3\right)}$, $\tau _3=\frac{2n}{2n+1}\left(c_s^2\tau
_2-\frac{v_tP1v_1}{2nv_t}\left(1-nc_s^2\right)-c_s^2v_5v_t\frac{P1}{2}\right)-\frac{P1'}{2}$, \\
$\tau _4=\frac{2n}{2n+1}\frac{v_tu}{1-u^2}\left(\frac{v_1}{nv_t}\left(1-nc_s^2\right)+c_s^2v_5\right)$, $v_1=\frac{\Delta'}{2\Delta
}+\frac{2}{r}-\left(2\lambda ^2v_t-a^2\right)v_t\frac{\text{P1}}{4F}$, \\
$D_3=\frac{u^2v_t\left(2\lambda^2v_t-a^2\right)}{2F}$, $D_4=\frac{1}{v_t}+\frac{2\lambda^2}{2\lambda^2v_t-a^2}-\frac{2\lambda^2v_t-a^2}{F}$,
$D_5=D_3\left(\frac{2}{u}+\frac{D_4v_tu}{1-u^2}\right)$, $\tau_2=\tau _1-\frac{v_t\left(2\lambda^2v_t-a^2\right)}{4F}P1'$, $v_5=\left(2\lambda^2v_t-a^2\right)\frac{P1}{4F}v_4$, \\
$\tau_1=\frac{1}{2}\left(\frac{\Delta''}{\Delta}-\frac{(\Delta')^2}{\Delta ^2}\right)-\frac{2}{r^2}$,
$v_4=\frac{v_3}{\left(2\lambda^2v_t-a^2\right)F}$, $v_3=\left(4\lambda^2v_t-a^2\right)F-\left(2\lambda^2v_t-a^2\right)^2v_t$.  

\subsection{Parameter space}

\noindent
Having presented the complete scheme of drawing the phase-space portrait numerically, we
focus our attention on the parameter space of the system. $\mathcal{E}$ is scaled by rest-mass energy
and includes both rest-mass energy and thermal energy components. Setting $\mathcal{E} = 1$ corresponds
to an initial state where no thermal energy is present. Furthermore, setting $\mathcal{E} < 1$
corresponds to initial conditions with negative energy. In this case, a dissipative mechanism
is needed to extract energy from the flow so that a flow solution is obtained with positive
energy. For our system of inviscid flow, this is not possible and thus we must consider flows
with $\mathcal{E} > 1$. All values of $\mathcal{E}$ greater than $2$, although possible, correspond to extremely high
initial thermal energy. Since this is not a common feature of accreting black hole systems,
it is usual to restrict the system within the parameter range $1 < \mathcal{E} < 2$. \\

\noindent
$\lambda = 0$ implies a spherically symmetric flow, where $\lambda > 4$ (in $G = M_{BH} = c = 1$ scaling) implies
that the flow is not anymore in the Keplerian regime. In this region, multi-critical solutions
do not generally occur. Thus we restrict ourselves to the parameter range of $0 < \lambda < 4$. \\

\noindent
In isothermal fluids, polytropic index $\gamma = 1$. $\gamma > 1$ corresponds to extremely dense fluids
where comparatively large magnetic fields with direction dependence, i.e, anisotropic pressure
are present. As we are not considering general relativistic magneto-hydrodynamics, we should
constrain ourselves in the domain $1 < \gamma < 2$. Moreover, throughout literature, the realistic
limits to polytropic index for accretion astrophysics is $\gamma = \frac{4}{3}$ for ultra-relativistic flows and
$\gamma = \frac{5}{3}$ for non-relativistic flows. Thus we will limit ourselves in the parametric range between
$\frac{4}{3} < \gamma < \frac{5}{3}$ . \\

\noindent
Here we mention pro-grade flows, i.e. where the flow co-rotates with black hole and retrograde
flows, i.e. where the flow counter-rotates with black hole. We consider both these
flows and in order to distinguish between the two, we allow positive and negative values of
$a$, whereas only positive values of $\lambda$ are allowed. Thus the range of $a$ is $-1 < a < 1$. An
upper limit of $0.998$ of a has been set in some literature where interaction with accretor and
the accreting material has been considered (\citep{thorne74apj}). In his work, Thorne considered the interaction of the accretion flow with the hole in such a way, that the accretion flow can alter the mass and spin of the hole, which, however, we do not consider in our present work. Thus, the present system of polytropic fluid accretion is studied within the parameter range
$\left[1<\mathcal{E}<2,0<\lambda<4,\frac{4}{3}<\gamma<\frac{5}{3},-1<a<1\right]$. \\

\begin{figure}[!htbp]
\centering
\includegraphics[scale=0.8]{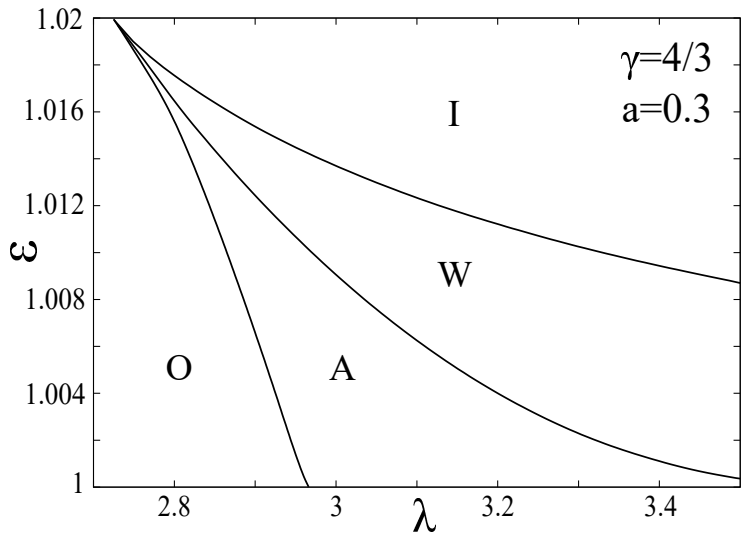}
\caption{$\mathcal{E}$ vs. $\lambda$ plot for polytropic NT disc with $a = 0.3$ and $\gamma = 4/3$}
\label{fig1}
\end{figure}

\noindent
Fig. (\ref{fig1}) depicts the characteristic parameter space diagram for a polytropic NT
disc. The NT disc has been selected for the purpose of demonstration because
it is the oldest prescription of hydrostatic equilibrium discs available in literature. All other
prescriptions display the same general properties in this regard. For a fixed set of $\left[\gamma, a\right]$,
possible multi-critical solutions form a wedge-shaped projection on the $\mathcal{E}-\lambda$ plane. The
multi-critical solutions constitute a set of three critical points, viz. $r_c^{in}$, $r_c^{mid}$ and $r_c^{out}$, such
that $r_c^{in}<r_c^{mid}<r_c^{out}$. The region $A$ represents the `accretion solutions' for which the
entropy accretion rate, $\dot{\Xi}(r_c^{in})>\dot{\Xi}(r_c^{in})$. The region $W$ consists of those solutions for which
$\dot{\Xi}(r_c^{in})>\dot{\Xi}(r_c^{in})$. Such solutions are known as `wind solutions'. The curve dividing regions 
$A$ and $W$ covers those critical points through which \textit{heteroclinic orbits} are formed in phase-space.
Slight perturbations in the flow due to turbulence or other physical factors can push
such solutions into either accretion or wind regime. Regions outside the wedge ($O$ and $I$)
contain mono-critical solutions. Inside region $O$, the critical point is of outer-type, which
means it forms far away from the horizon, whereas inside region $I$, the critical points are
formed nearer to the horizon and are known as inner-type. Both regions $O$ and $I$ contain
single critical points (corresponding to mono-transonic accretion/wind) upto a certain limit
of flow parameters beyond which critical solutions cease to exist. However, since we are
interested only in the $A$ region, a detailed discussion regarding the relation between system
parameters and the existence or non-existence of critical points lies beyond the scope of the
present article.

\subsection{General relativistic polytropic shock conditions}

\noindent
Since, we have assumed a non-dissipative, inviscid flow, the specific energy and mass accretion
rate are conserved. Thus, shocks formed in such flows must also preserve the conserved
quantities. We consider the shock surface to be infinitesimally thin such that there are
no temperature gradients within shock leading to any unwanted dissipation. Hence the
discontinuity must satisfy the general relativistic Rankine-Hugoniot conditions (\citep{eckart40pr}, \citep{taub48pr}, \citep{lichnerowicz67}, \citep{thorne73apj}, \citep{taub78anrfm}, \citep{hacyan82grg}, \citep{abd06cqg}) given below.

\begin{equation}
\begin{aligned}
&[[\rho v^\mu \eta_\mu]]=[[\rho v^r]]=0\\
&[[T_{t\mu}\eta^\mu]] = [[(p+\varepsilon)v_{t}v^r]]=0\\
&[ [T_{\mu\nu}\eta^\mu\eta^\nu]]=[[(p+\varepsilon)(v^r)^2+pg^{rr}]]=0
\end{aligned}
\end{equation}

\noindent
Where $ \eta_\mu=\delta^r_\mu $ is orthonormal to the surface of shock formation. For any arbitrary flow variable $f$, $ [[f]] $ is defined as $ [[f] ]= f_+-f_- $ , where $ f_+ $ and $ f_- $ are values of $ f $ just outside and inside the shock, respectively. The difference measures the discontinuity in the flow variable due to shock. The first condition is conservation of mass accretion rate and the other two conditions are energy-momentum conservation. These conditions must be satisfied at the location of shock formation. In order to find out the location of shock formation, a shock invariant quantity, which depends only on $ u,c_s $ and $ \gamma $, is constructed using the conditions above. The first and second conditions are trivially satisfied owing to the constancy of the mass accretion rate and the specific energy. The first condition is basically $ (\dot{M})_+ = (\dot{M})_- \ $ and third condition is $ (T^{rr})_+=(T^{rr})_- $. Thus a shock invariant quantity $S_{\rm sh}$ can be defined as

\begin{equation}
S_{\rm sh} = T^{rr}/ \dot{M}
\end{equation}
which also satisfies $ [[S_{\rm sh}]]=0 $. \\

\noindent
In order to calculate the shock invariant quantity we note that $ h $ corresponds to the enthalpy of the stationary solutions of the steady atate flow,  given by equation Eq. (\ref{enthalpy}). $ c_{s}=(1/h)dp/d\rho = (1/h)k\gamma \rho^{\gamma-1}$, which gives $ \rho $ (and hence also $ p $ and $ \epsilon $) in terms of $ k,\gamma $ and $ c_{s} $. Thus,

\begin{equation}
\begin{aligned}
& \rho = k^{-\frac{1}{\gamma-1}}\left[\frac{(\gamma-1)c_{s}^2}{\gamma(\gamma-1-c_{s}^2)}\right]^{\frac{1}{\gamma-1}}\\
& p = k^{-\frac{1}{\gamma-1}}\left[\frac{(\gamma-1)c_{s}^2}{\gamma(\gamma-1-c_{s}^2)}\right]^{\frac{\gamma}{\gamma-1}}\\
& \varepsilon = k^{-\frac{1}{\gamma-1}}\left[\frac{(\gamma-1)c_{s}^2}{\gamma(\gamma-1-c_{s}^2)}\right]^{\frac{1}{\gamma-1}}\left(1+\frac{c_{s}^2}{\gamma(\gamma-1-c_{s}^2)}\right)
\end{aligned}
\end{equation}

\noindent
Now $ \dot{M} = {\rm constant}\times rH(r) \rho v^r $ and $ T^{rr}=(p+\varepsilon)(v^r)^2+pg^{rr} $, where $ v^r = u/\sqrt{g_{rr}(1-u^2)}  $. \\

\paragraph{NT \& RH discs}
The shock-invariant quantity $ S_{\rm sh} = T^{rr}/\dot{M} $ becomes

\begin{equation}\label{shockinv_NT_poly}
S_{\rm sh} = \frac{(u^2(\gamma-c_s^2)+c_s^2)}{uc_s\sqrt{(1-u^2)(\gamma-1-c_s^2)}}
\end{equation}

\noindent
where we have removed any over all factor of $ r $ as shock invariant quantity is to be evaluated at  $ r=r_{\rm sh} $ for different branches of flow. \\

\paragraph{ALP discs}
In this case, the shock-invariant quantity turns out to be

\begin{equation}\label{shockinv_ALP_poly}
S_{\rm sh} = \frac{(u^2(\gamma-c_s^2)+c_s^2)\sqrt{\lambda^2 v_t^2 - a^2 (v_t -1)}}{uc_s\sqrt{(1-u^2)(\gamma-1-c_s^2)}}
\end{equation}

\noindent
where $v_t$ is given in (\ref{v_t}). \\

\begin{figure}[!htbp]
\begin{tabular}{cc}
\includegraphics[scale=0.8]{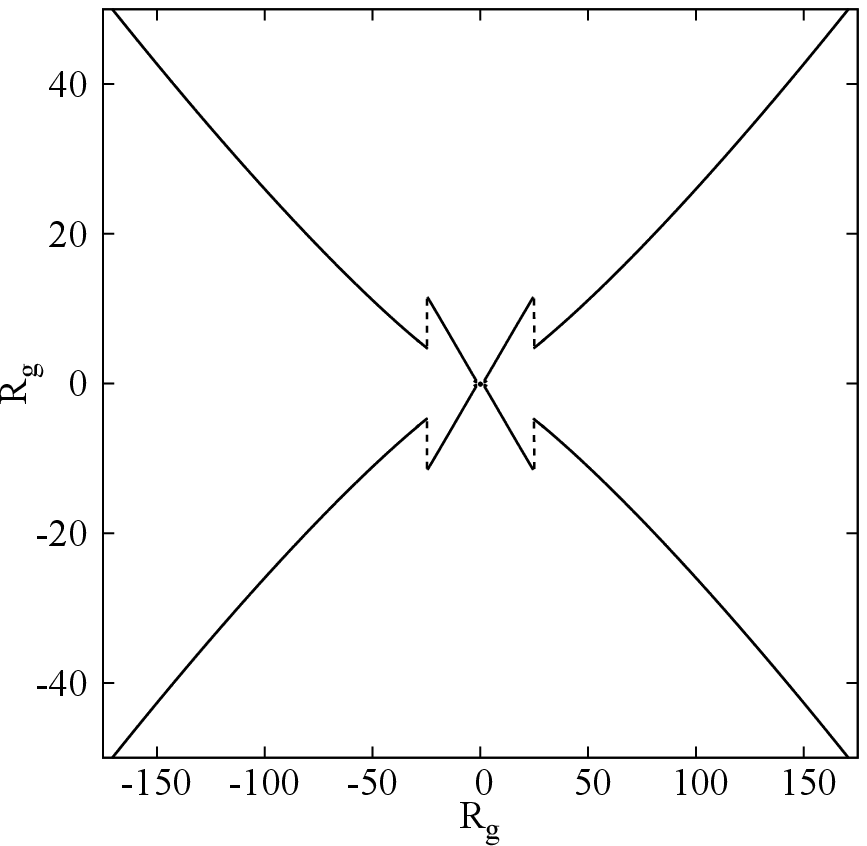} &
\includegraphics[scale=0.8]{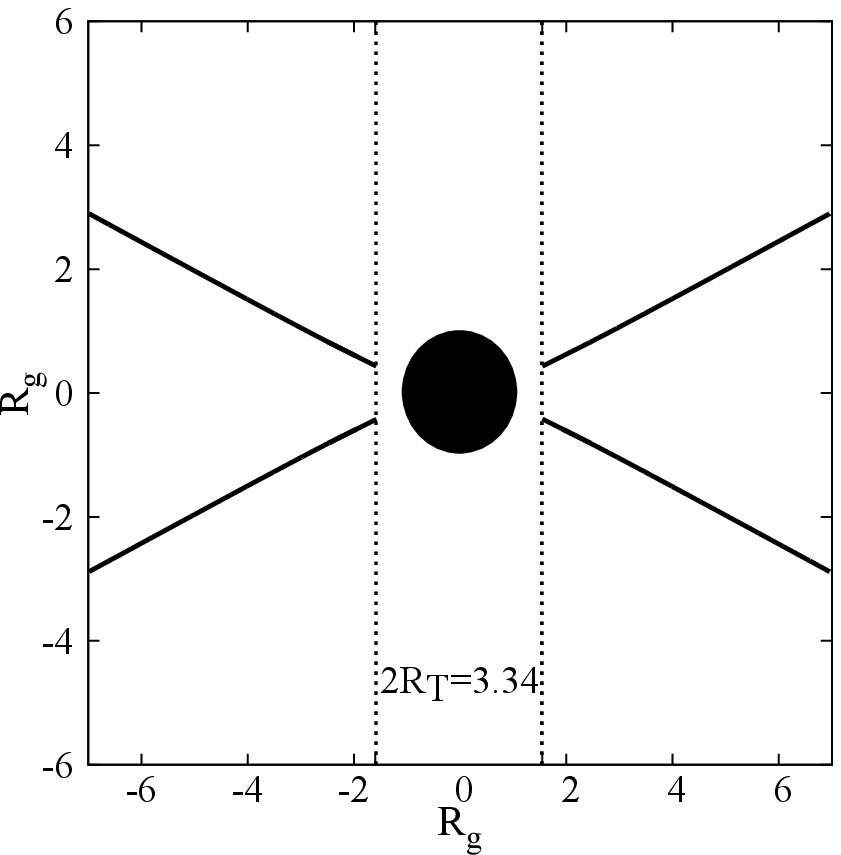} \\
\includegraphics[scale=0.8]{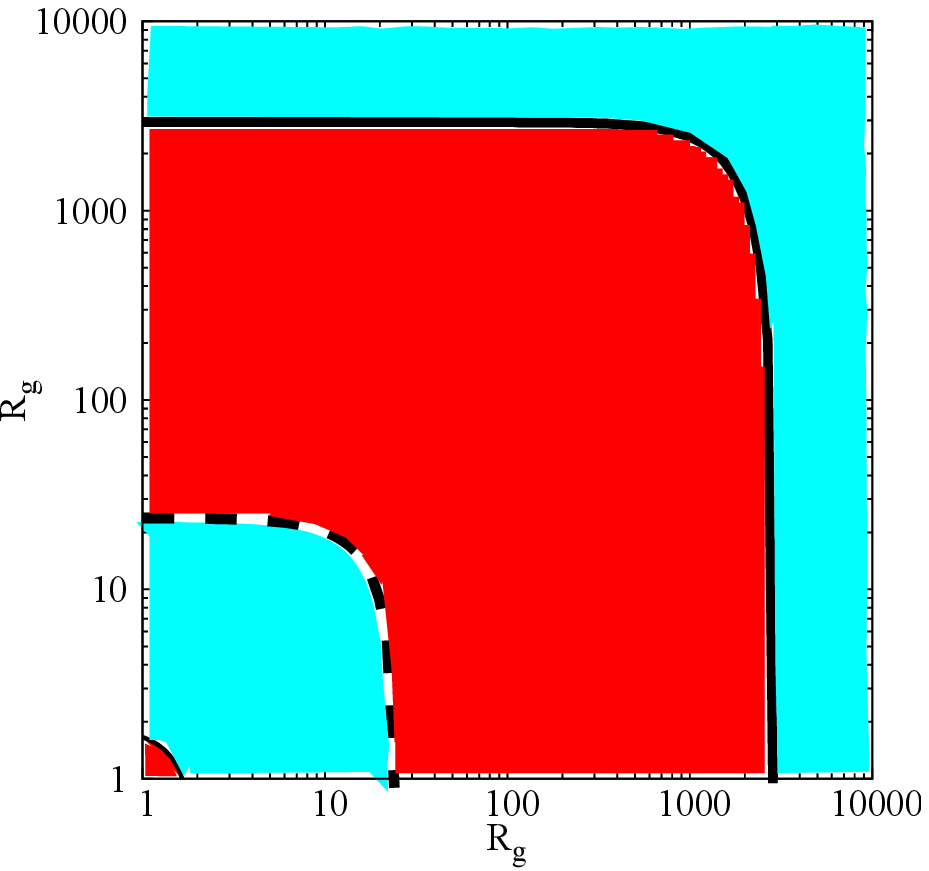} &
\includegraphics[scale=0.8]{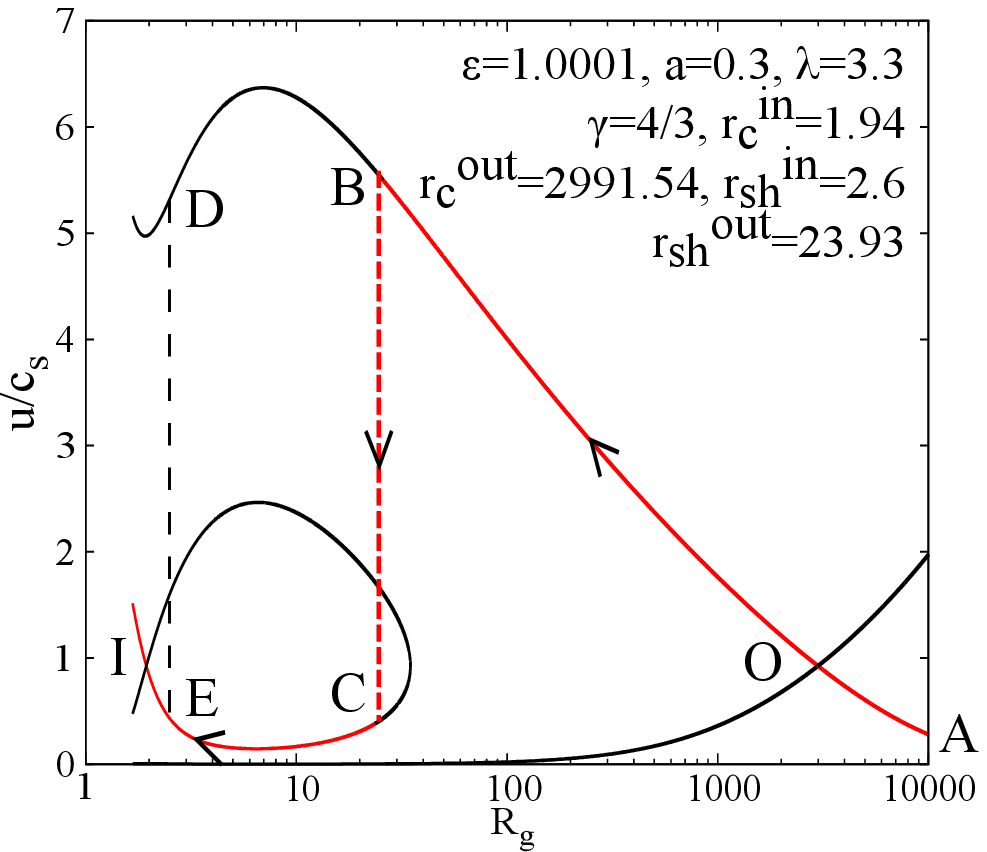} \\
\end{tabular}
\caption{(a) Disc height vs. radial distance, (b) Central region in (a) is magnified, (c)
Transonic boundaries - Cyan (lighter shade) region represents subsonic flow and red (darker
shade) region represents supersonic flow, (c) Phase space trajectories - Mach number vs.
radial distance}
\label{fig2}
\end{figure}

\noindent
Fig.\ref{fig2}(a) shows edge-on view of the polytropic NT disc for flow with a given
value of $\left[\mathcal{E},\lambda,\gamma,a\right]$ in the presence of shock. Since specific 
energy is conserved in polytropic
accretion, the post-shock flow sees a discontinuous increase in temperature, density and pressure.
Consequently, the disc gets `puffed-up' at the shock location as is evident from the
plot. A closer look at the central region of the disc (fig.\ref{fig2}(b)) reveals that 
the disc gets terminated
abruptly at a given radius, known as the `trauncation radius' ($r_T$). The artefact, as
discussed earlier, is due to the inherent mathematical limitation of the NT and
RH class of discs regarding their closest approach of the event horizon. It limits the
use of such discs for obtaining flow variables in close vicinity of the horizon (\textit{quasi-terminal
values}) that are essential for shadow-imaging of black holes (\citep{tbnd19prd}). 
Fig.(\ref{fig2}(c)) depicts face-on view
of one of the quadrants of the same disc on a logarithmic scale. The regions colored in cyan
(lighter shade) and red (darker shade) represent regions of subsonic and supersonic flows
respectively. The solid boundary curves lie over points of continuous transonicity (transonic
points corresponding to $r_c^{out}$ and $r_c^{in}$), while the dashed boundary curve 
lies over the points
of discontinuous transonicity, i.e. shock. The picture becomes clear in fig.(\ref{fig2}(d)) where a
complete phase-space profile ($u/c_s$ vs. $r$) for the given combination of flow parameters has
been presented. Trajectory of the physical flow (marked in red) in the presence of shock
has been indicated with arrows. The flow starts subsonically through point $A$ and proceeds 
to cross the first transonic point $O$ (corresponding to the outer critical point $r_c^{out}$) beyond
which it attains supersonic velocities till point $B$. Here the flow encounters shock, causing
a discrete jump onto point $C$ on the \textit{homoclinic orbit}. The shock (dashed line $BC$) lies
at the location $r_{sh}^{out}$, which can be calculated by looking for those values of $r$ where $S_{sh}$ on the upper and lower branches become equal in magnitude. Using this procedure, a second
shock location ($r_{sh}^{in}$, shown with the black dashed line $DE$) is sometimes obtained such that
$r_{sh}^{in} < r_{sh}^{out}$ . But such inner second shocks have been found to be unstable 
in previous works (\textbf{references}). In the absence of shock, the flow would have continued 
supersonically along the upper branch through $B$, effectively resulting in mono-transonic accretion. 
However, transition onto $C$ brings the flow down to subsonic regime, and subsequently it follows the
trajectory of the lower branch through the second transonic point $I$ (corresponding to the
the inner critical point $r_c^{in}$) and proceeds beyond to fall into the horizon. \\

\begin{figure}[!htbp]
\begin{tabular}{cc}
\includegraphics[scale=0.8]{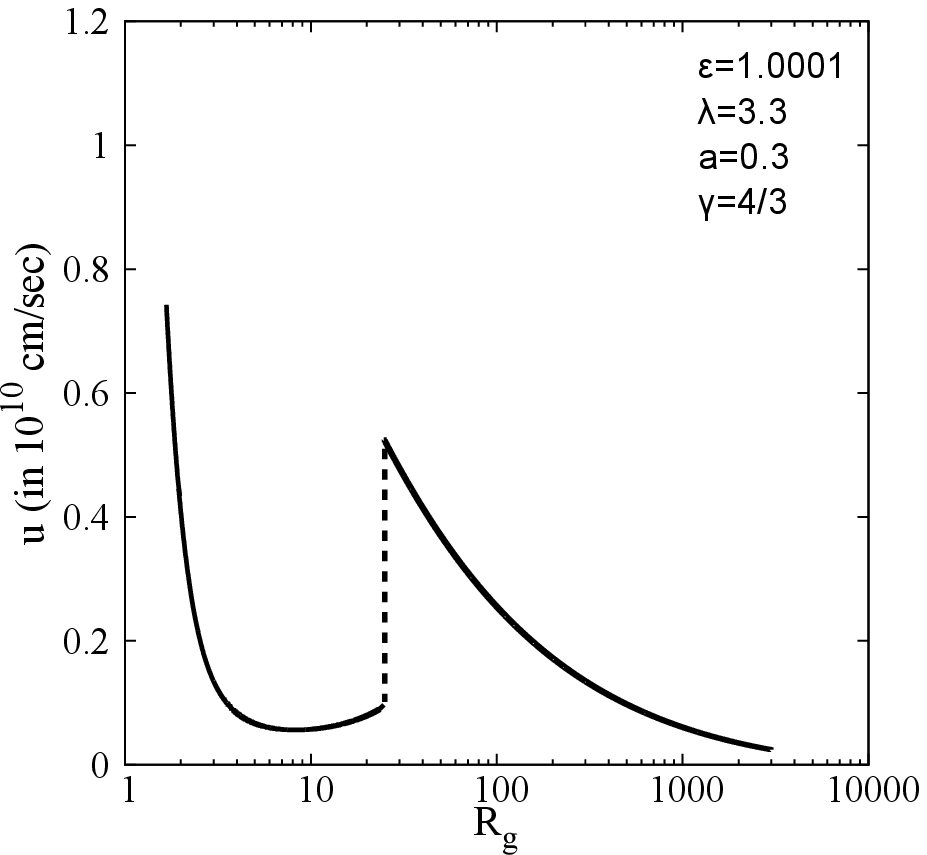} &
\includegraphics[scale=0.8]{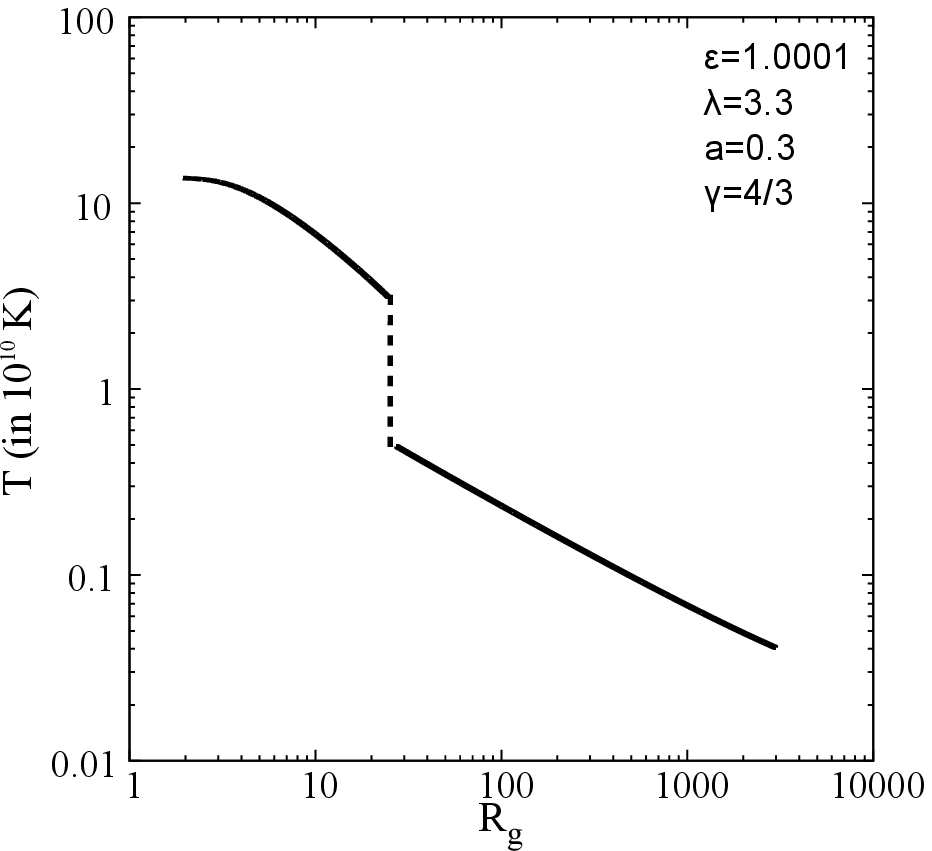} \\
\includegraphics[scale=0.8]{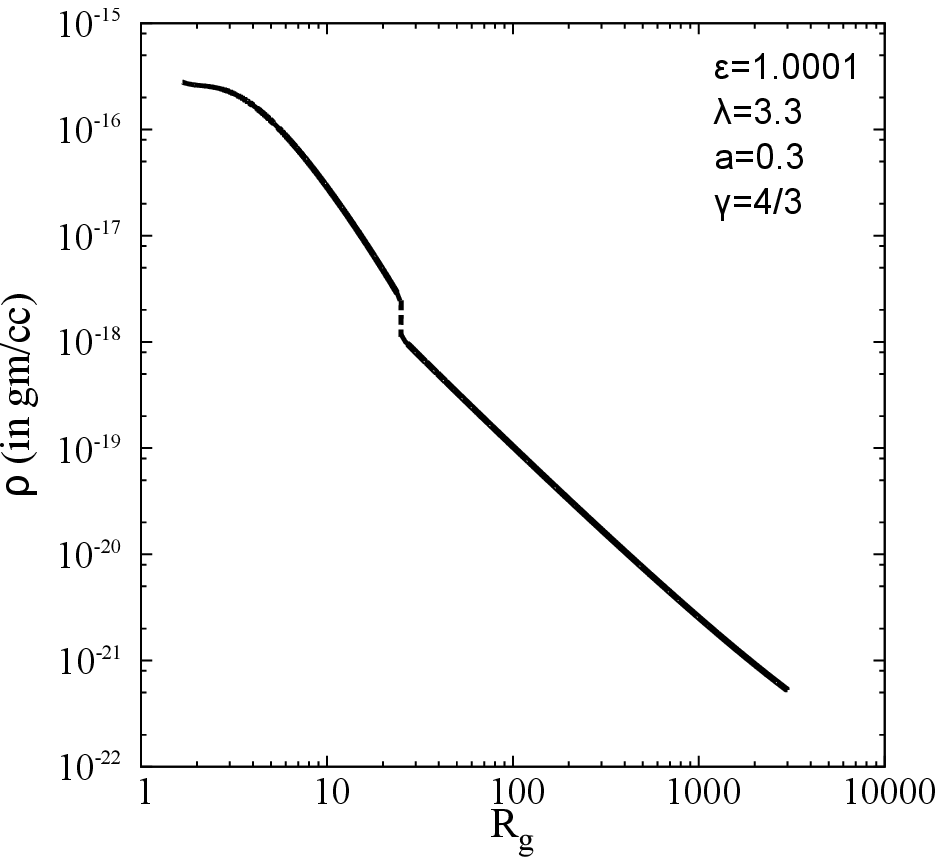} &
\includegraphics[scale=0.8]{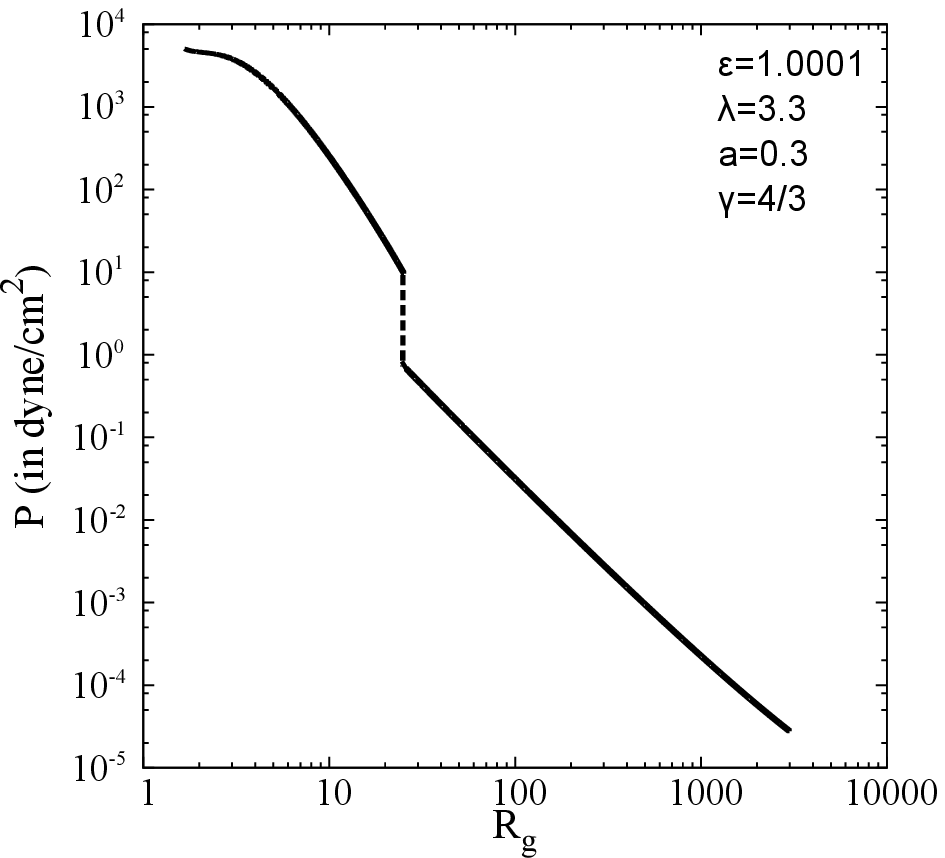} \\
\end{tabular}
\caption{Generic flow profiles - (a) Advective velocity ($u$) vs. $R_g$, (b) Flow ion temperature
($T$) vs. $R_g$, (c) Rest mass density ($\rho$) vs. $R_g$ (d) Pressure ($P$) vs. $R_g$. 
$u$ is in units of $10^{10}$ cm/sec, $T$ is in units of $10^{10}$ Kelvin, $\rho$ is in units of gm/cc 
and $P$ is in units of dyne/cm$^2$.}
\label{fig3}
\end{figure}

\noindent
In fig.(\ref{fig3}), we plot the variation of the flow velocity $u$ (fig.(\ref{fig3}(a)), 
flow temperature $T$ (fig.(\ref{fig3}(b)), matter density $\rho$ (fig.(\ref{fig3}(c)) 
and the fluid pressure $P$ (fig.(\ref{fig3}(d)) as a function of the radial distance
as measured from the horizon in terms of the Schwarzschild radius $R_g$ ($=2GM_{BH}/c^2$).
In fig.(\ref{fig3}(a)-(d)), the variation is shown as a combination of two solid lines connected by
a vertical dashed line. The solid line at the right of the dashed vertical line represents the
variation along the flow solution passing through the outer sonic point (starting point of the
solid line) and ending at the shock. The dashed vertical line corresponds to the discontinuous
jump of the physical variable ($u$, $T$ etc.) at the shock location. The solid line to the left
of the dashed vertical line represents the variation along the integral flow solution starting
from the shock location and ending at the corresponding truncation radius for NT/RF discs
given by eqn. (\ref{r_truncated})). It is evident from the figure that the accretion flow 
slows down at the shock and gets compressed. Such relatively slow, shock-compressed post-shock 
flow becomes hotter and denser. The energy-preserving hotter flow adiabatically expands and hence the
post-shock part of the disc gets puffed-up, as explained earlier. \\

\noindent
As mentioned in introduction, our motivation is to compare various astrophysical properties
of the shocked flow for three different disc thicknesses. We thus need to find out the region
of the parameter space (parameters for which the shock forms) common to all such three
different flow thicknesses. \\

\begin{figure}[!htbp]
\begin{tabular}{cc}
\includegraphics[scale=0.8]{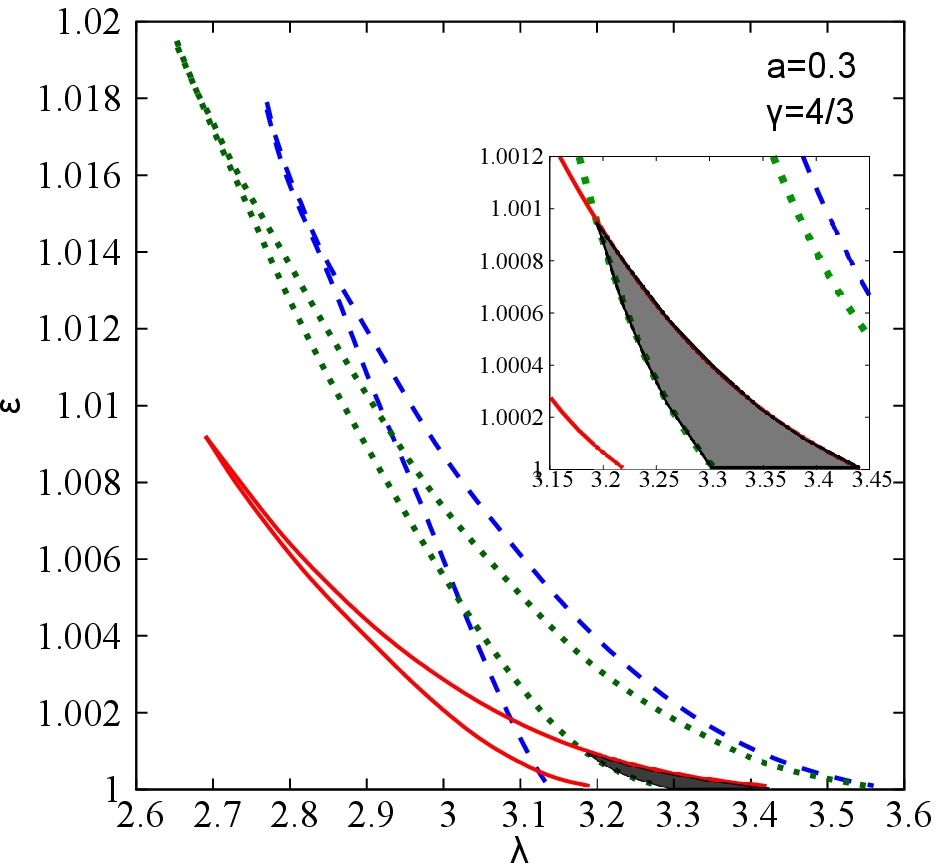} &
\includegraphics[scale=0.8]{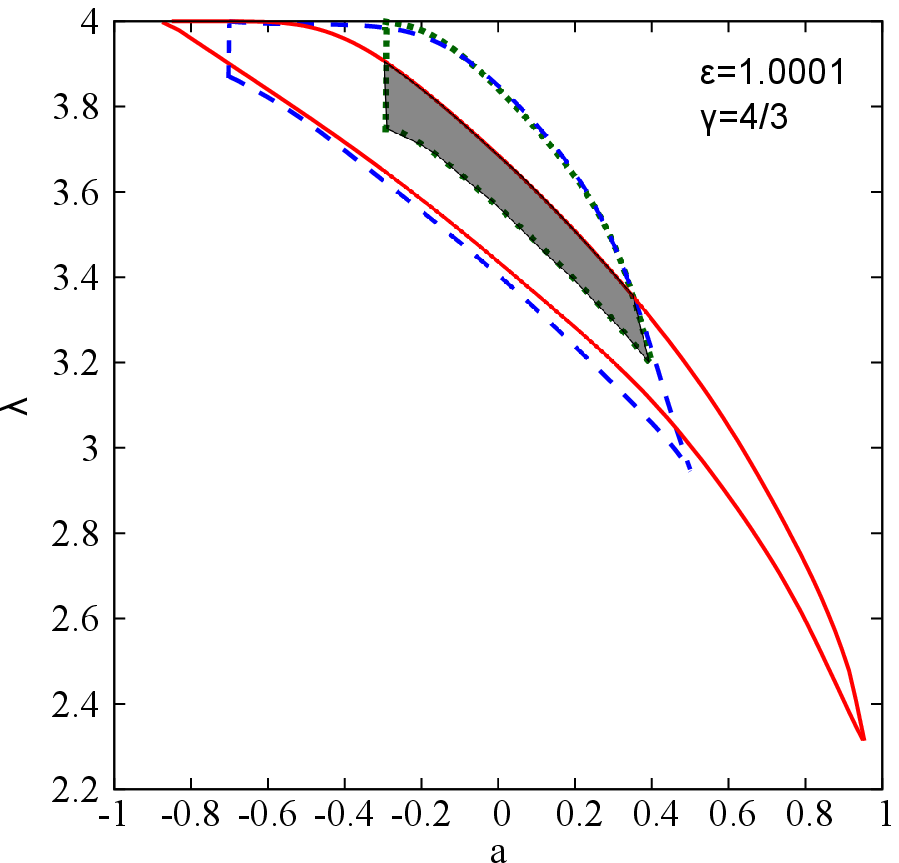} \\
\end{tabular}
\caption{Parameter space overlap for shocked solution - (a) $\mathcal{E}-\lambda$ plot and 
(b) $\lambda-a$ plot. Insets and shaded areas depict the common regions of shock solutions. 
ALP shown by red solid lines, RH shown by green dotted lines and NT shown by blue dashed lines.}
\label{fig4}
\end{figure}

\noindent
In fig.(\ref{fig4}(a)), we plot the $\left[\mathcal{E},\lambda\right]$ regions for which 
the shock forms for the flow having thickness
as prescribed by NT (blue dashed curve in the online version of this article),
RH (green dotted curve in the online version) and ALP
(red solid curve in the online version). It is to be mentioned that from now onwards, the line
types (solid, dotted and dashed) and the line colors (red, green and blue) corresponding to
the three different disc models (ALP, RH and NT respectively) will be used in the same order
(as used in the present diagram) throughout the paper, be it for polytropic or isothermal flow. \\

\noindent
In fig.(\ref{fig4}(a)), the overlap of the parameter spaces for the shock-forming flow corresponding 
to three different disc models is shown using dark-grey shade. The grey shaded common
region has also been demonstrated in the inset of the figure. The figure has been drawn by
keeping the values of the black hole spin and the polytropic index of the flow to be fixed.
The values of such fixed parameters are shown in the figure. Such values are representative
values only, i.e. the shocked flow can be obtained for other set of values of $\left[a,\gamma\right]$ as well. \\

\noindent
Fig.(\ref{fig4}(b)) shows the parameter space diagram spanned by the flow angular momentum and
the spin angular momentum of the black hole for a fixed set of values of 
$\left[\mathcal{E},\gamma\right]$ as specified in
the figure. The set of values of $\left[\mathcal{E},\gamma\right]$ is representative and 
similar $\left[\lambda,a\right]$ space can be obtained
for other values of $\left[\mathcal{E},\gamma\right]$ as well. We choose the particular set of values 
$\left[\mathcal{E}=1.0001,\gamma=4/3\right]$
so that we can cover an extended range of the Kerr parameter to identify the shocked solutions.
It is evident from the figure that the lower values of the black hole spin allow shock
formation for relatively larger values of flow angular momentum, as well as for a relatively
large span of values of the angular momentum. This is probably obvious because a lower
spin accretor effectively reduces the influence of the flow angular momentum. \\

\begin{figure}[!htbp]
\begin{tabular}{cc}
\includegraphics[scale=0.8]{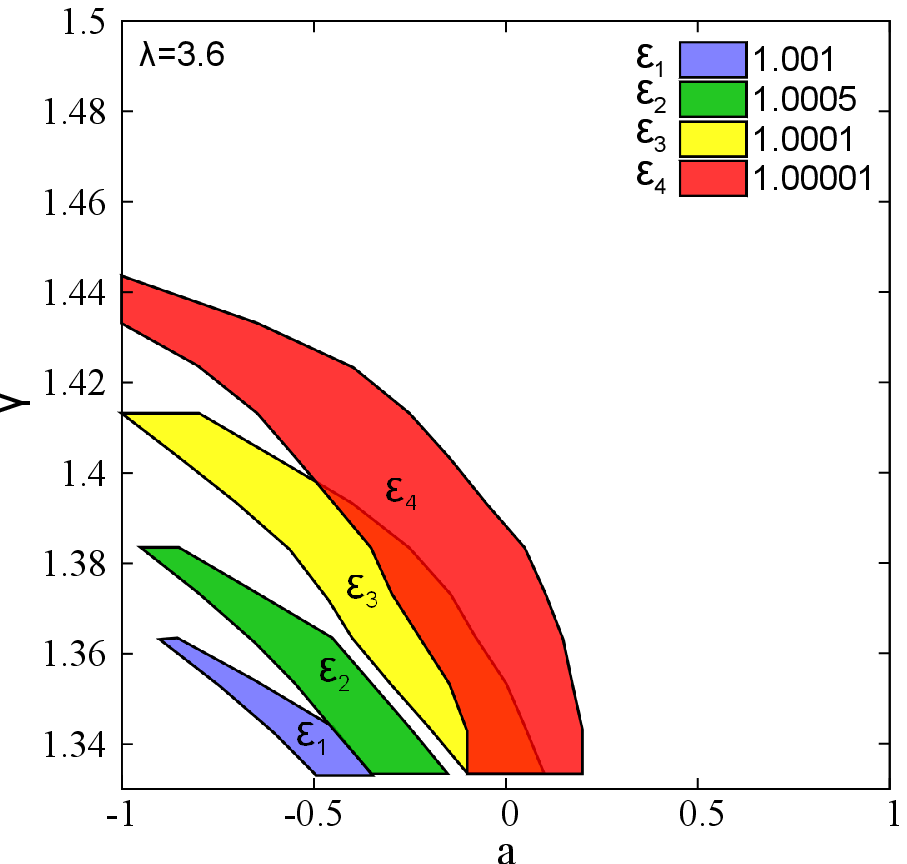} &
\includegraphics[scale=0.8]{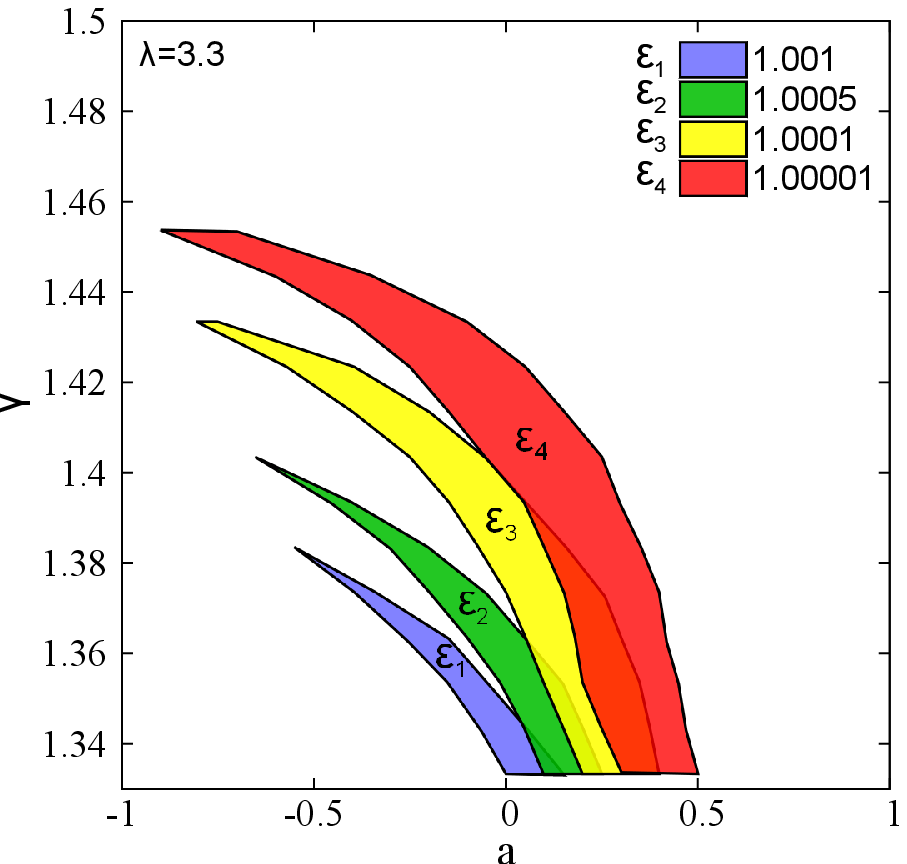} \\
\includegraphics[scale=0.8]{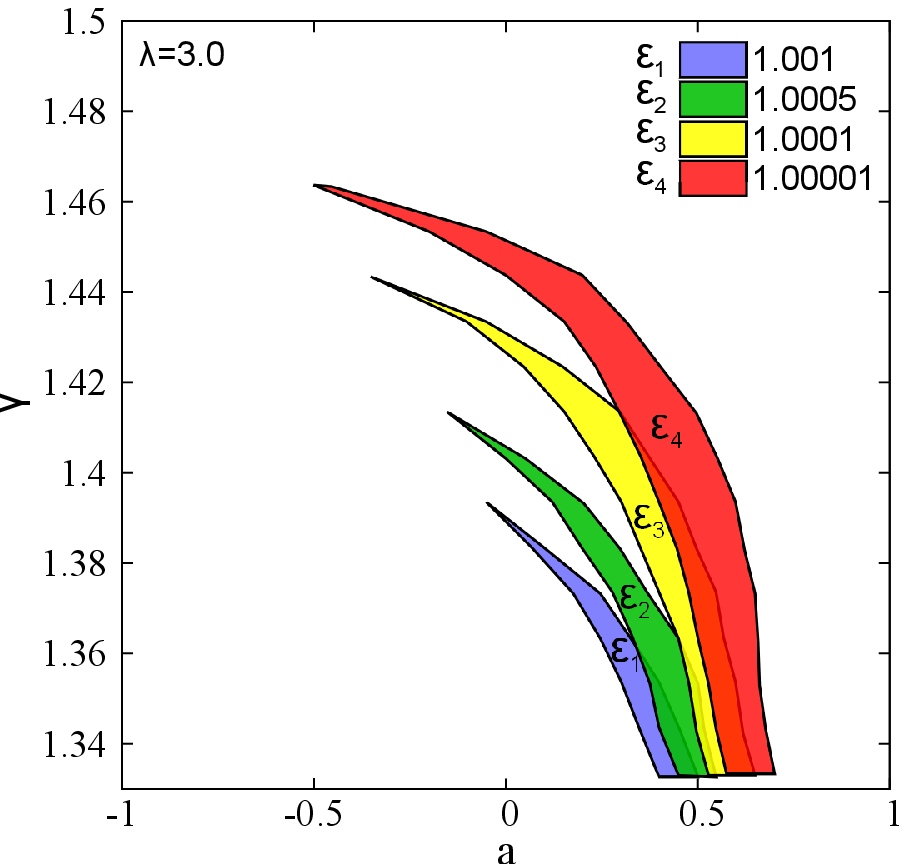} &
\includegraphics[scale=0.8]{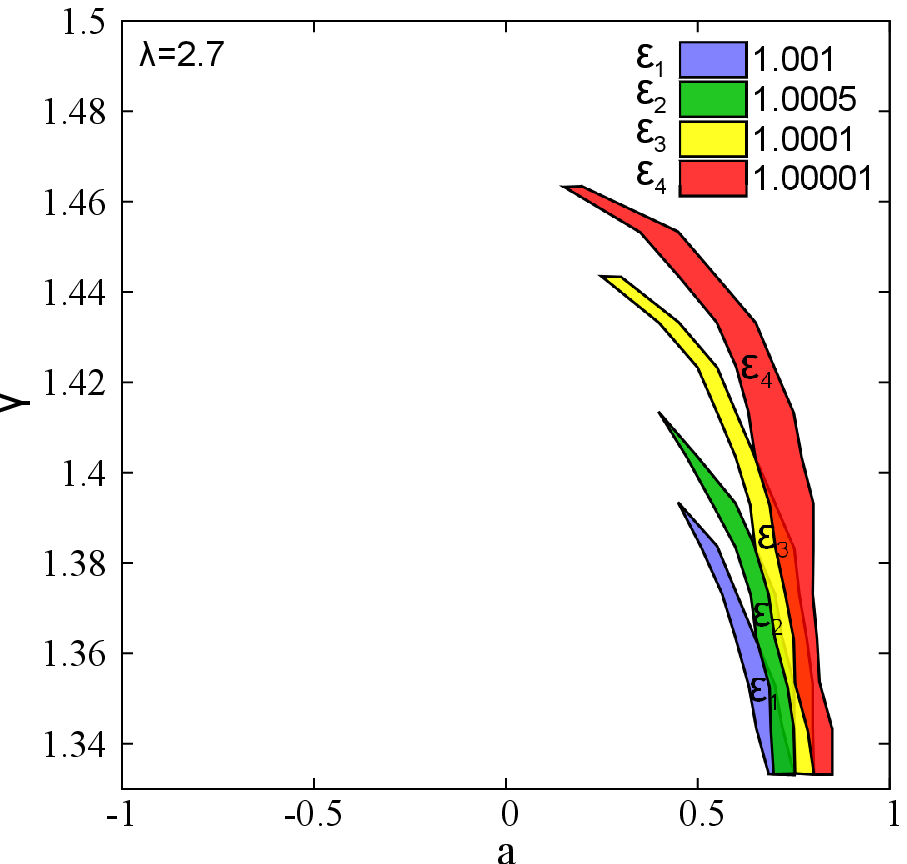} \\
\end{tabular}
\caption{$\gamma$-$a$ plot with different $\mathcal{E}$ values ($\mathcal{E}_1=1.001$, 
$\mathcal{E}_2=1.0005$, $\mathcal{E}_3=1.0001$, $\mathcal{E}_4=1.00001$) for (a) $\lambda=3.6$, 
(b) $\lambda=3.3$, (c) $\lambda=3.0$ and (d) $\lambda=2.7$.}
\label{fig5}
\end{figure}

\noindent
We conclude our discussions on parameter dependence of the shock solutions by studying
the role of adiabatic index $\gamma$ and specific energy $\mathcal{E}$. 
Fig.(\ref{fig5}) depicts $\left[\gamma,a\right]$ space with shock
solutions for different values of $\mathcal{E}$ and $\lambda$ corresponding to the 
ALP-type of discs. Similar
panels can be constructed for NT and RH discs as well, but the trends of variation have
been found to be similar. Due to the quality of not being constrained with any truncation
radius and thus providing the maximum scope to look for shocks in terms of radial distance,
the ALP disc has been chosen for the purpose of demonstration in this regard. \\

\noindent
In fig.(\ref{fig5}(a)), we see that the relevant adiabatic indices anti-correlate with the black hole
spin parameter. Flows with four different values of specific energy ($\mathcal{E}_1 > \mathcal{E}_2 > \mathcal{E}_3 > \mathcal{E}_4$,
values provided in the respective figures and marked with blue, green, yellow and red colours
respectively) have been studied. We find that flows with lower values of $\mathcal{E}$ can 
lead to formation of shocks over a greater range of $\gamma$ from the fully relativistic limit of 
$\gamma=4/3$ till
other intermediate values below the non-relativistic limit of $\gamma=5/3$. The lowest value of
$\mathcal{E}$ considered here ($\mathcal{E}_4=1.00001$) serves our purpose of explanation. However, even lower
values of specific energy can be considered to predict shock solutions theoretically almost
over the entire astrophysically relevant range of $\gamma$ (from fully relativistic to non-relativistic
flows). This, of course, comes with an obvious trade-off between the spans of results achieved
and the computational costs incurred. \\

\noindent
Figs.(\ref{fig5}(b)-(d)) shows similar $\gamma$ vs. $a$ plots with for parameters at which shocks form. The four separate figures (a)-(d) in the panel indicate the successively decreasing values of the flow angular
momentum $\lambda$. We have already shown in fig.(\ref{fig4}(b)) that the flow angular momentum and
black hole spin anti-correlate with each other in the context of shock formation. Hence it is
expected that as the value of $\lambda$ is decreased for a given set of $\left[\mathcal{E},\gamma\right]$, shocks will be obtained
at higher values of $a$. That is exactly what we see along figs.(\ref{fig5}(b)-(d)). However, it
should also be noted that the range of $a$ over which such solutions are obtained decreases
significantly with decreasing flow angular momentum. The inter-relationships between all
the system parameters are extremely complex for such highly non-linear systems. A definitive
picture can only be procured through generation of a complete 4-dimensional parameter 
space diagram. Scanning the entire possible parameter space is heavily time-consuming and
computationally exhausting, and hence is beyond our present theoretical scope. However,
an integrated study of various parameter combinations as presented in our work, provides a
sufficiently comprehensive assessment of the relevant shock regimes. \\

\begin{figure}[!htbp]
\begin{tabular}{cc}
\includegraphics[scale=0.7]{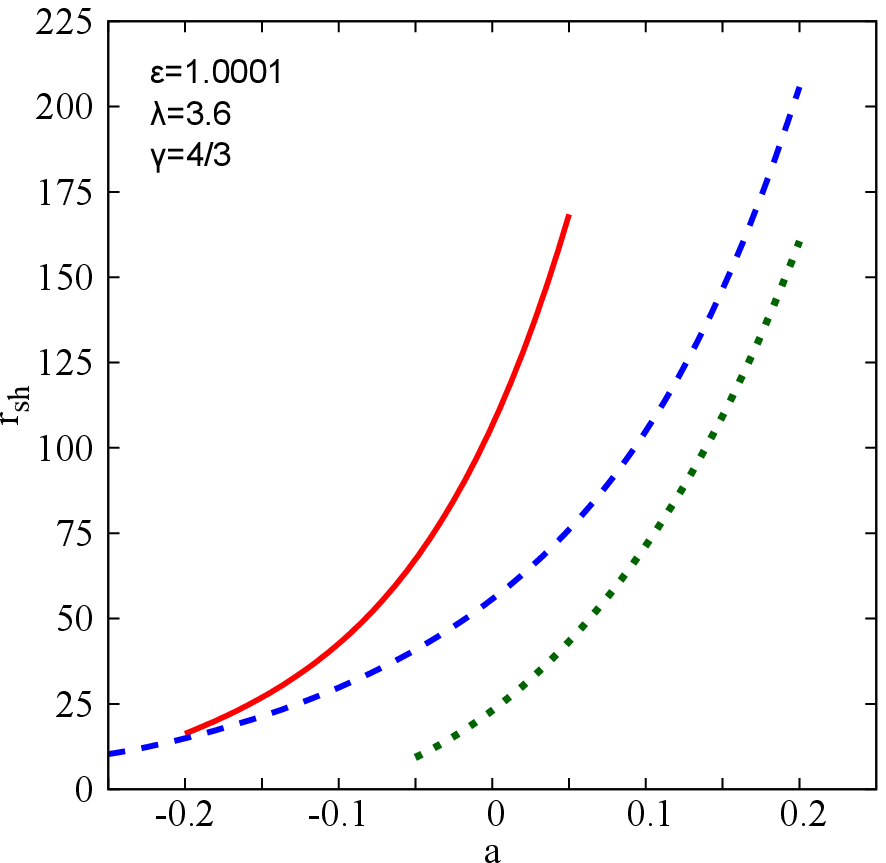} &
\includegraphics[scale=0.7]{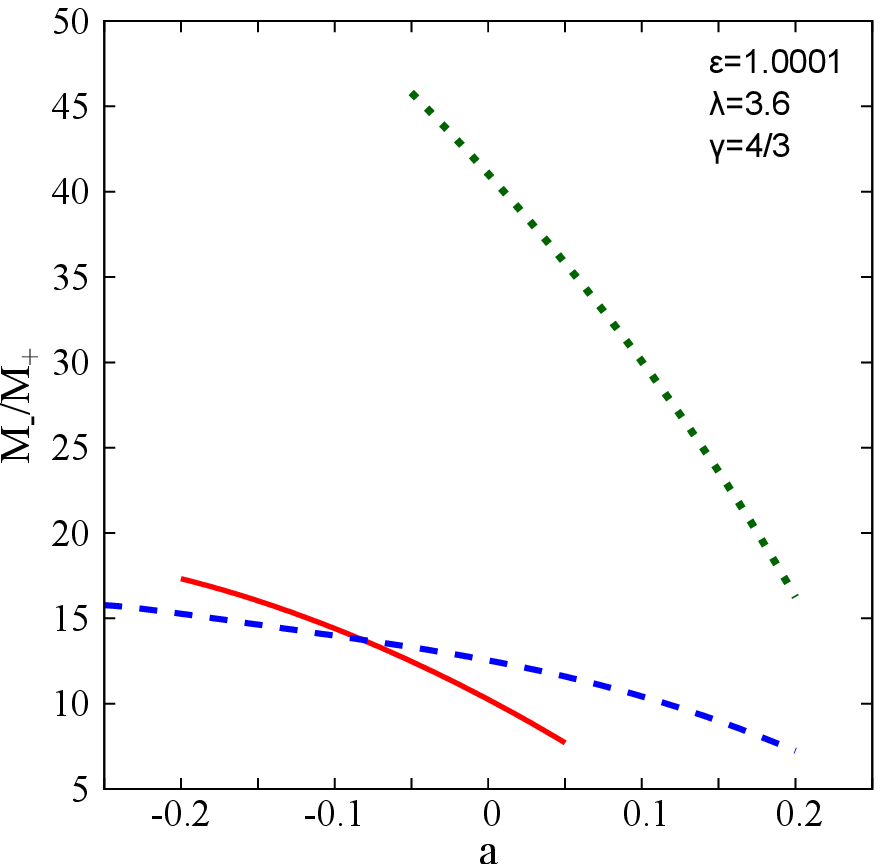} \\
\includegraphics[scale=0.7]{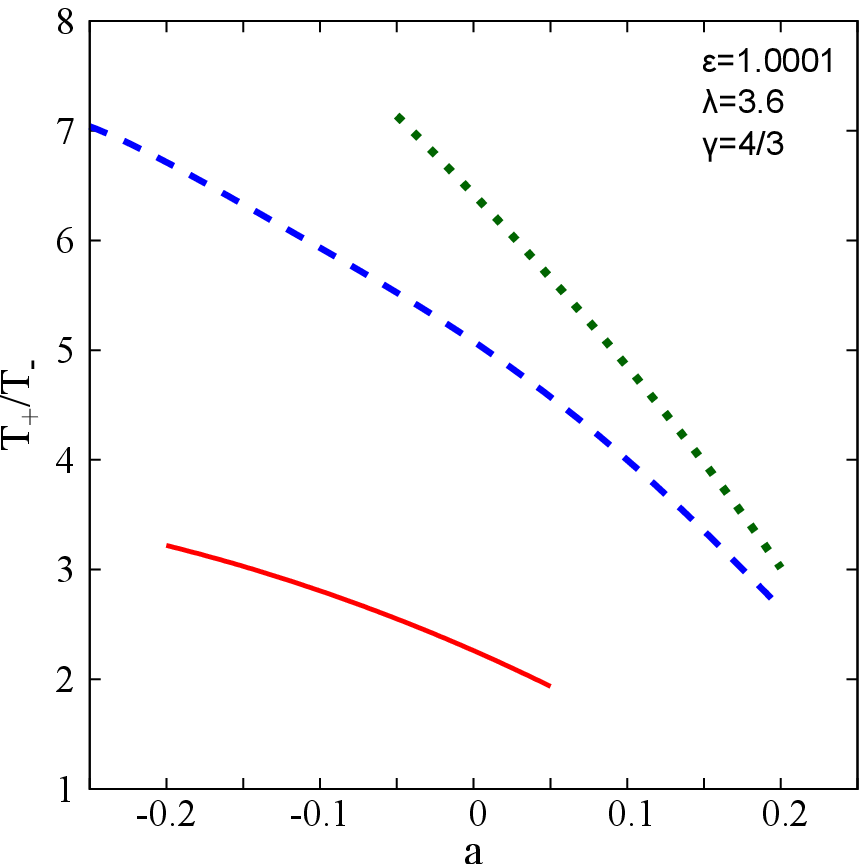} &
\includegraphics[scale=0.7]{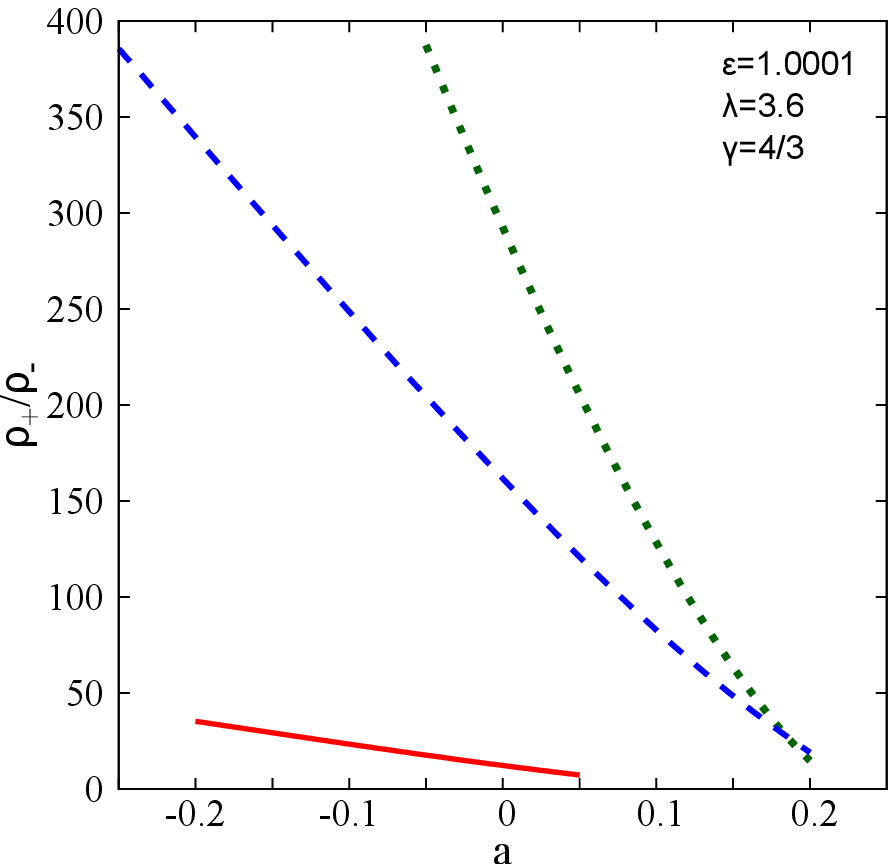} \\
\includegraphics[scale=0.7]{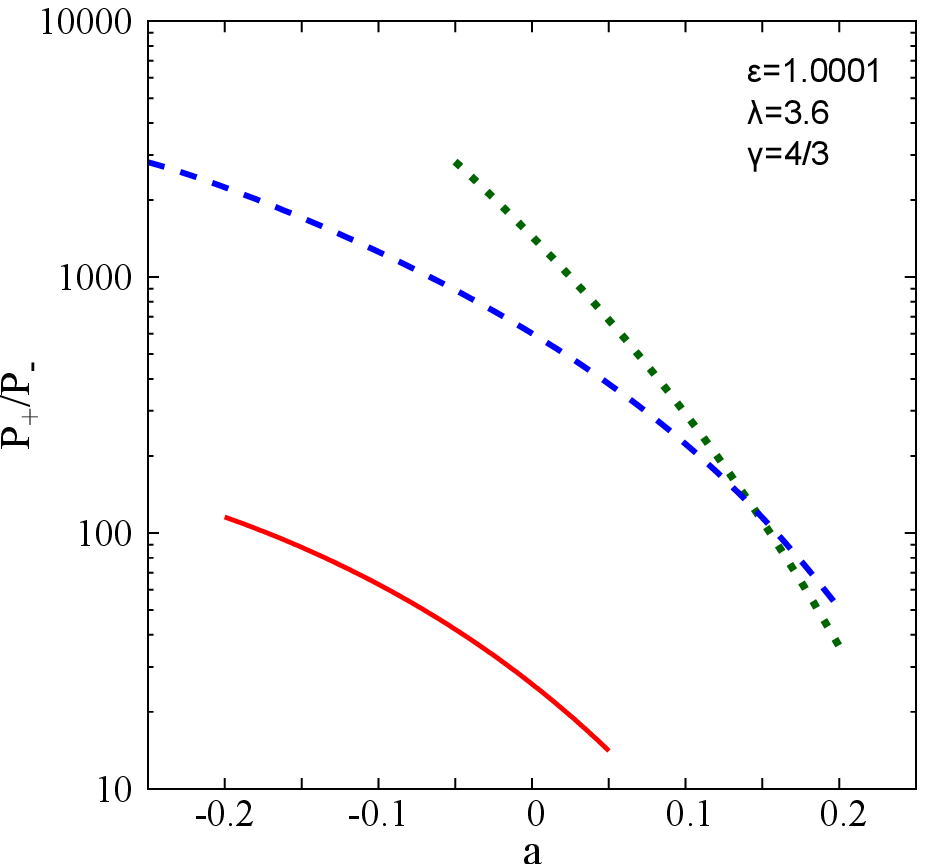} &
\includegraphics[scale=0.7]{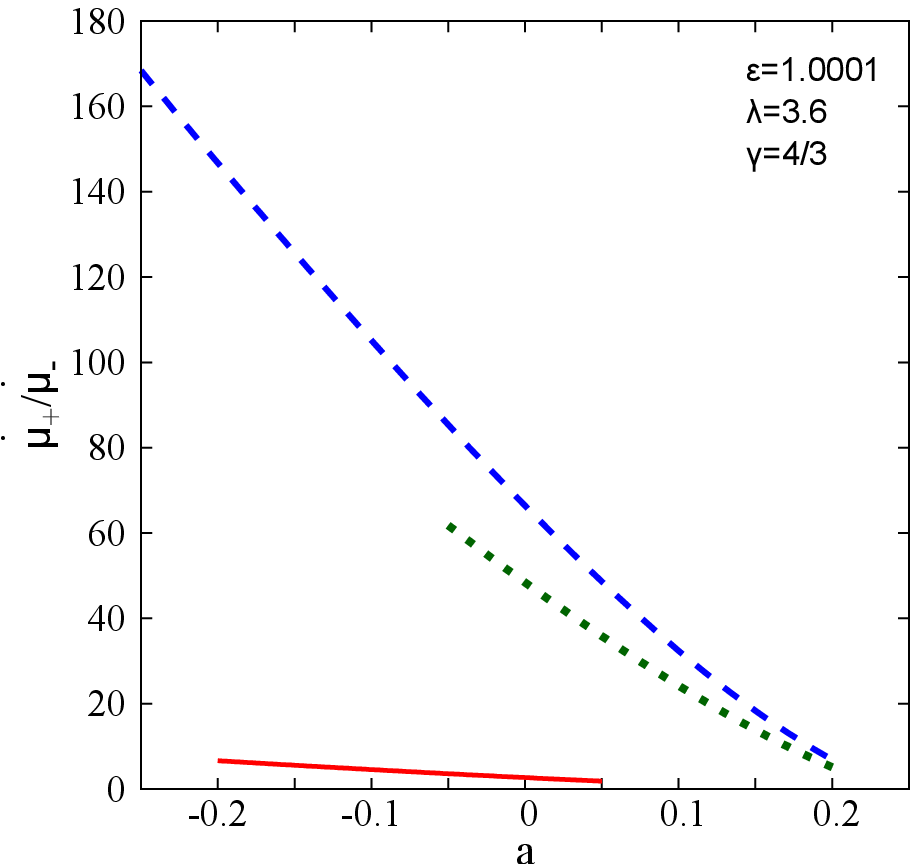} \\
\end{tabular}
\caption{Shock location - (a) $r_{sh}$ (in terms of $R_g$) vs. $a$, Ratios at shock - (b) $M_-/M_+$ vs.
$a$, (c) $T_+/T_-$ vs. $a$, (d) $\rho_+/\rho_-$ vs. $a$, (e) $P_+/P_-$ vs. $a$ and (f) $\dot{\Xi}_+/\dot{\Xi}_-$ vs. $a$. `$-$' and `$+$' refer
to values `before' and `after' the shock respectively. ALP shown by red solid lines, RH shown by green
dotted lines and NT shown by blue dashed lines.}
\label{fig6}
\end{figure}

\noindent
In fig.(\ref{fig6}(a)), we plot the shock locations (measured from the horizon in units of the 
Schwarzsc-\ hild radius $R_g$ ($=2GM_{BH}/c^2$)), and other shock-related quantities as a function 
of the black hole spin for three different disc models. The set of values of 
$\left[\mathcal{E},\lambda,\gamma\right]$ are kept fixed, and their fixed values are shown in 
the respective figures. \\

\noindent
We observe that the shock location ($r_{sh}$) co-relates with the spin parameter of the black
hole. This is intuitively obvious because higher spin effectively enhances the effect of flow
angular momentum. Greater the angular momentum, larger will be the distance at which the
centrifugal barrier forms. The shock under consideration is centrifugal pressure-supported.
Hence, $r_{sh}$ is pushed farther away from the horizon with increasing values of $a$. For fixed
values of $\left[\mathcal{E},\lambda,\gamma,a\right]$ the shock forms farthest for ALP-type of disc, 
whereas it forms closest for RH discs. For NT-type discs, the shock forms at an intermediate distance. 
$r_{sh}$ vs. $a$ curve for ALP-type discs approaches that for NT-type discs asymptotically but they never
intersect. This has been investigated for values of $\left[\mathcal{E},\lambda,\gamma,a\right]$ other 
than those used to generate fig.(\ref{fig6}). With decreasing $\lambda$, the overall set of $r_{sh}$-$a$ curves shift towards higher values
along the $a$ and $r_{sh}$ axes. Thus, we find that $r_{sh}$ anti-correlates with $\lambda$ as expected since $r_{sh}$
co-relates with $a$ and $a$ anti-correlates with $\lambda$. Similarly, the authors have verified that $r_{sh}$
anti-correlates with $\gamma$ and co-relates with $\mathcal{E}$ (since $\gamma$ anti-correlates and 
$\mathcal{E}$ co-relates with $a$,
as shown in fig.(\ref{fig5})). \\

\noindent
At the shock location, directed flow velocity gets randomized and the gravitational potential
energy available at the shock location determines the shock strength. The closer the shock
forms to the horizon, the stronger it should be. Hence the strength should anti-correlate
with the shock location and thus with the black hole spin parameter. This is exactly what
we observe in fig.(\ref{fig6}(b)). The shock strength is defined as the ratio of the pre- to post-shock
Mach number of the flow. We plot the shock strength ($M_-/M_+$, hereafter any accretion
variable with a subscript `-' would indicate that it has been measured at the shock location
before the shock is formed, i.e. it has been measured on the integral solution passing
through the outer sonic point, and variables with subscript `+' would refer to the post-shock
values measured at the shock location on the integral solution passing through the inner
sonic point) as a function of the Kerr parameter for both prograde and retrograde flow. As
argued above, we clearly see that the shock strength anti-correlates with the black hole spin
for both co-rotating as well as counter-rotating flows. We observe an intersection of the $M_-/M_+$
vs. $a$ curve for the ALP and NT disc models. Such intersection, by any means, does not
indicate any degeneracy in the disc models, i.e. it does not mean that for certain values
of $\left[\mathcal{E},\lambda,\gamma,a\right]$, two or more separate disc models provide 
the same value of any significant
accretion variable. It is important to note that the ratio of Mach numbers can assume same
values at the point of intersection, but not the value of any individual quantity. The ratio
of Mach numbers can be the same for two (or more) different set of post-shock values. The
shock strength is found to co-relate with $\lambda$ and $\gamma$, and anti-correlate with 
$\mathcal{E}$ as expected from
the relation between the corresponding parameters with a shown previously in the respective
parameter-space diagrams. In fig.(\ref{fig6}(c)) and (\ref{fig6}(d)), the post- to pre-shock 
temperature and shock compression ratio (ratio of flow densities after and before the shock) 
have been plotted
against the change of black hole spin. As expected, these quantities anti-correlate with $a$,
because greater the amount of available gravitational potential energy at the shock, higher
will be the amount of temperature changes and larger will be the amount of compression.
It is evident from the figure that the RH-type of discs become most dense and hot after the
shock forms. Whereas, the ALP-type of discs change their temperature and density in minimum
amounts at the shock. The NT kind of flow assumes an intermediate value for these
two ratios. More or less, similar trends are observed for the variation of the ratio of the post
to the pre-shock fluid pressure for three different disc models. Here too, we find intersection
among the two curves, but as explained earlier, it does not indicate any type of degeneracy.
Finally, in fig.(\ref{fig6}(f)) we plot the ratio of the post to pre-shock entropy accretion rates for
three disc models as a function of the black hole spin. The ratio of the entropy accretion
rate is a measure of entropy production at the shock. As we observe, such measure may not
have any one-to-one correspondence with the shock strength. The entropy is directly related
to the expression of the mass accretion rate of the steady-state flow. \\

\noindent
In passing, we would like to mention that the set of $\left[\mathcal{E},\lambda,\gamma\right]$ used 
to draw this figure is not
unique by any means. We chose this set of values just to have a reasonable span of the black
hole spin covering both prograde as well as retrograde flows. It is to be noted that shock
does form for accreting black holes with intermediate as well as higher spin, for both co- and
counter-rotating flows. One can obtain shocked flows for high-spin accretors using a suitable
set of $\left[\mathcal{E},\lambda,\gamma\right]$.

\section{Isothermal accretion}

\subsection{Fluid equations}

\noindent
The equation of state characterising isothermal fluid flow
is given by,

\begin{equation}\label{eqtn_of_state_isothermal}
p=c_s^2\rho=\frac{\cal R}{\mu}\rho T=\frac{k_B\rho T}{\mu m_H}
\end{equation}

\noindent
where $T$ is the bulk ion temperature, $\cal R$ is the universal gas constant, $k_B$ is Boltzmann constant, $m_H$ is mass of the Hydrogen atom and $\mu$ is the mean molecular mass of fully ionized hydrogen. The temperature $T$ as introduced in the above equation, and which has been used as one of the parameters to describe the isothermal accretion, is the temperature-equivalent of the bulk ion flow velocity. That is the reason why the value appears to be high ($10^{10}-10^{11}$ K) in this work. The actual disc temperature is the corresponding electron temperature, which should be of the of the order of $10^6-10^7$ Kelvin. Now using the equation of state (\ref{eqtn_of_state_isothermal}), the equations needed to draw the phase portrait will be derived.

\subsubsection{Energy-momentum equation}

\noindent
Using eqn. (\ref{eqtn_of_state_isothermal}), eqn. (\ref{Euler}) can be rewritten as

\begin{equation}\label{omegamunu}
u^\nu \left[D_\nu(\rho^{c_s^2} u_\mu) - D_\mu(\rho^{c_s^2} u_\nu) \right] = 0
\end{equation}

\noindent
Using the time component of this equation and the stationary nature of the flow one obtains the conserved quantity

\begin{equation}\label{conserved_isothermal}
\xi=v_t \rho^{c_s^2}.
\end{equation}

\noindent
Taking the logarithmic derivative of eqn. (\ref{conserved_isothermal}), the derivative of density $\rho^{'}$ is obtained as

\begin{equation}\label{density_derivative}
\frac{\rho^{'}}{\rho} = \frac{u^{'}}{u(u^2 - 1)} - \left(\frac{f^{'}}{f} + \frac{\Delta^{'}}{2\Delta} \right)
\end{equation}

\subsubsection{Continuity equation}

\noindent
In this section we again derive the velocity gradient for two separate classes, one consisting of NT and RH and the other consisting of the ALP height prescription. We note that we can still integrate continuity equation and the conserved quantity mass accretion rate $ \dot{M} $ as defined in eqn. (\ref{Psi}). \\

\paragraph{NT \& RH discs}

\noindent
Using the fact that $\frac{p}{\rho}$ is the constant $c_s^2$, the height of the disc for these two height prescriptions in case of isothermal accretion can be written as

    \begin{equation}\label{isothermal_height_NT}
    H(r) = c_s f(r,a)    
    \end{equation}

    Using eqn. (\ref{isothermal_height_NT}) and putting the value of $\frac{\rho^{'}}{\rho}$ in the logarithmic derivative of eqn. (\ref{Psi}), we obtain

    \begin{equation}\label{dudr_NT_iso}
    \frac{du}{dr} = \frac{u(1-u^2)\left[c_{s}^2(\frac{\Delta'}{2\Delta}+\frac{f'}{f})+\frac{1}{2}(\frac{B'}{B}-\frac{\Delta'}{\Delta})\right]}{u^2-c_s^2}=\frac{N}{D}.
    \end{equation}

\noindent
    Again we mention that for NT height recipe, $f(r,a)$ is replaced by $f_{NT}(r,a)$ as defined in eqn. (\ref{f_NT}) and for RH height recipe, we replace $f(r,a)$ by $f_{RH}(r,a)$ as defined in eqn. (\ref{RHhieght}). \\

\paragraph{ALP discs}

\noindent
    For this recipe, the height function in the case of isothermal acretion is

    \begin{equation}\label{isothermal_height_RH}
    H(r) = c_s^2 \sqrt{\frac{2r^4}{\lambda^2 v_t^2 - a^2 (v_t - 1)}}.
    \end{equation}

\noindent
    Following the same procedure as used in previous class of height recipes, one yields

\begin{equation}\label{dudr_ALP_iso}
\frac{du}{dr}=\frac{{c_s}_c^2(\frac{\Delta'}{2\Delta}+\frac{2}{r}-(2\lambda^2 v_t-a^2)\frac{v_t P}{4F})-\frac{P}{2}}{\frac{u}{1-u^2}-\frac{{c_s}_c^2}{u(1-u^2)}(1-(2\lambda^2 v_t-a^2)\frac{u^2 v_t}{2F})}
\end{equation}

\noindent
where $P=\frac{\Delta'}{\Delta } - \frac{B'}{B}$ and $F=\lambda^2 v_t^2-a^2(v_t-1)$.

\subsection{Critical point conditions}

\noindent
Follwing the same scheme as in polytropic process, we find the slopes of directrices at critical points, solve for the radial position of critical point, $r_c$ and draw the phase portrait. Again we present NT and RH-type of discs in the first class and ALP-type of discs in the next class for reasons stated earlier.

\subsubsection{NT \& RH discs}

\noindent
Setting $ D = 0 $ in Eq. (\ref{dudr_NT_iso}) yields

\begin{equation}\label{critpoints_NT_iso}
u^2|_c =c_{s}^2 |_c
\end{equation}

\noindent
Setting $N=0$ yields

\begin{equation}\label{c_s_critical_iso}
c_{s}^2|_c =
\frac{\frac{\Delta'}{\Delta}-\frac{B'}{B}}{(\frac{\Delta'}{2\Delta}+\frac{f'}{f})}.
\end{equation}

\noindent
To find the critical points for sothermal accretion the method followed is different from that of polytropic accretion as the basic parameter characterising the flow is different for polytropic and isothermal accretion. In polytropic accretion, the parameters are $\mathcal{E}$ and $\lambda$, wheras the isothermal flow is characterized by the parameters $T$ and $\lambda$. So, by putting the chosen value of the temparature $T$ in Eq. (\ref{eqtn_of_state_isothermal})  to find the costant sound speed. The the value of $c_s^2$ is used in Eq. (\ref{c_s_critical_iso}) and the equation

\begin{equation}\label{rc_NT_iso}
\frac{\cal R}{\mu} T = \frac{\frac{\Delta'}{\Delta}-\frac{B'}{B}}{(\frac{\Delta'}{2\Delta}+\frac{f'}{f})}
\end{equation}

\noindent
The right hand side of eq. (\ref{rc_NT_iso}) is function of the variable $r$ and by solvng this equation the critical points are obtained. \\

\noindent
The two values of the slopes at critical point is obtained from the quadratic equation

\begin{equation}\label{twoslopes_NT_iso}
\alpha_1 \left( \frac{du}{dr} \right)^2 - \alpha_2 \left( \frac{du}{dr} \right) - \alpha_3 = 0
\end{equation}

\noindent
where,
\begin{equation}
\alpha_1 = 4u_c
\end{equation}

\begin{equation}
\alpha_2 = \left(\frac{B'}{B} - \frac{\Delta'}{\Delta} + \left( \frac{\Delta'}{\Delta} + \frac{2f'}{f} \right)u^2_c \right)\left( 1-3u^2_c \right)
\end{equation}

\begin{equation}
\alpha_3 = u_c\left( 1-u^2_c \right) \left[ 2u^2_c \left( \frac{\Delta''}{2\Delta} + \frac{f^{''}}{f} -\frac{1}{2}\left(\frac{\Delta'}{\Delta}\right)^2 -\frac{f'}{f}^2 \right) - \frac{\Delta''}{\Delta} + \frac{B^{''}}{B} +\left(\frac{\Delta'}{\Delta}\right)^2 -\left(\frac{B'}{B}\right)^2 \right]
\end{equation}

\noindent
Thus we are equipped with all the information needed to draw the phase portrait diagram for a given parameter set of $\left[T, \lambda, a\right]$.

\subsubsection{ALP discs}

\noindent
The critical point conditions obtained by setting $N=0$ and $D=0$ are: \\

\begin{eqnarray}
u_c^2|_{VE}=\frac{P1}{\frac{\Delta'}{\Delta}+\frac{4}{r}} \\
{c_s}_c^2|_{VE}=\frac{u_c^2}{1-\frac{u_c^2v_t(2\lambda^2 v_t-a^2)}{2F}}
\label{cs_ALP_iso}
\end{eqnarray}

\noindent
Velocity gradient at critical points: \\
\begin{equation}
\left(\frac{du}{dr}\right)_c|_{VE}=-\frac{\beta_{VE}}{2\alpha_{VE}} \pm \frac{1}{2\alpha_{VE}}\sqrt{\beta_{VE}^2-4\alpha_{VE}\Gamma_{VE}}
\label{eqn73}
\end{equation}

\noindent
where, \\
$\alpha_{VE}=\frac{1+u_c^2}{\left(1-u_c^2\right)^2}-D_2D_6$, $\beta_{VE}=D_2D_7+\tau_4$, $\Gamma_{VE}=-\tau_3$, \\
$D_2=\frac{c_s^2}{u\left(1-u^2\right)}\left(1-D_3\right)$, $D_6=\frac{3u^2-1}{u\left(1-u^2\right)}-\frac{D_5}{1-D_3}$, \\
$D_7=\frac{D_3D_4v_tP1}{2\left(1-D_3\right)}$, $\tau _3=\left(c_s^2\tau_2-c_s^2v_5v_t\frac{P1}{2}\right)-\frac{P1'}{2}$, \\
$\tau _4=\frac{c_s^2v_5v_tu}{1-u^2}$, $v_1=\frac{\Delta'}{2\Delta
}+\frac{2}{r}-\left(2\lambda ^2v_t-a^2\right)v_t\frac{\text{P1}}{4F}$, \\
$D_3=\frac{u^2v_t\left(2\lambda^2v_t-a^2\right)}{2F}$, $D_4=\frac{1}{v_t}+\frac{2\lambda^2}{2\lambda^2v_t-a^2}-\frac{2\lambda^2v_t-a^2}{F}$,
$D_5=D_3\left(\frac{2}{u}+\frac{D_4v_tu}{1-u^2}\right)$, $\tau_2=\tau _1-\frac{v_t\left(2\lambda^2v_t-a^2\right)}{4F}P1'$, $v_5=\left(2\lambda^2v_t-a^2\right)\frac{P1}{4F}v_4$, \\
$\tau_1=\frac{1}{2}\left(\frac{\Delta''}{\Delta}-\frac{(\Delta')^2}{\Delta ^2}\right)-\frac{2}{r^2}$,
$v_4=\frac{v_3}{\left(2\lambda^2v_t-a^2\right)F}$, $v_3=\left(4\lambda^2v_t-a^2\right)F-\left(2\lambda^2v_t-a^2\right)^2v_t$. \\

\noindent
The location of critical points are solved just as described before by putting appropriate $T$ in eq. (\ref{eqtn_of_state_isothermal}) and then solving eq. (\ref{cs_ALP_iso}) by using the corresponding value of $c_s$.

\subsection{Parameter space}

\begin{figure}[!htbp]
\centering
\includegraphics{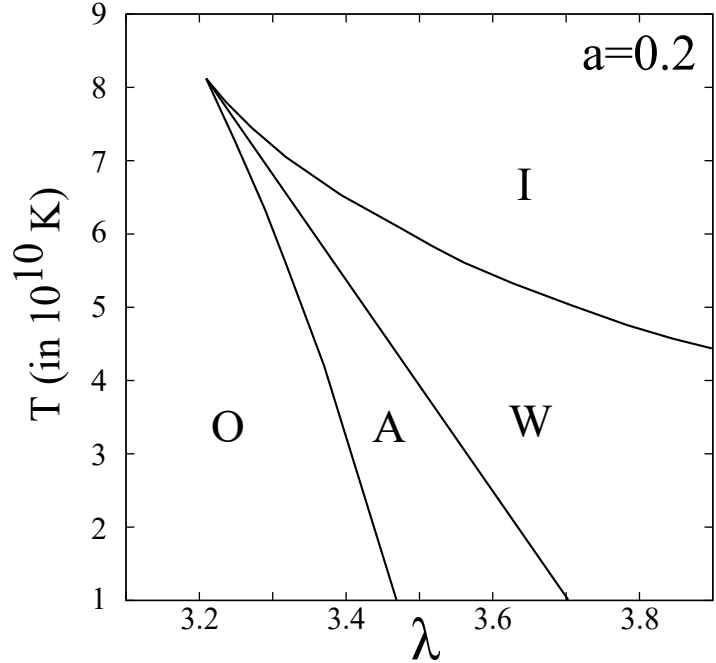}
\caption{$T-\lambda$ parameter space plot for accretion and wind in isothermal NT
disc at $a=0.2$}
\label{fig7}
\end{figure}

\noindent
In fig.(\ref{fig7}), we show that the parameter space spanned by the (constant) bulk flow temperature
and the flow angular momentum for a particular value of the black hole spin ($a = 0.2$).
Similar diagrams can be produced for other values of $a$, both for prograde as well as retrograde
flows. As discussed in section $3.3$, the parameter space is divided into four different
regions, $O$, $I$, $A$ and $W$ as shown in the figure. The parameter space has been constructed
for discs with NT-type of flow thickness. \\

\noindent
Both regions $O$ and $I$ produce a single sonic (critical) point. For $O$, the sonic point is outertype,
i.e. it forms far away from the horizon. Whereas for $I$, it is inner-type, i.e. it forms very
close to the horizon. Parameter space region marked by $A$ designates accretion flow with
three critical points. If shock forms, then the largest (outermost) and smallest (innermost)
critical points may become sonic points and two different integral accretion solutions, passing
through the outermost and the innermost critical (sonic) points respectively, may be joined
using a stationary shock solution. For flow characterised by parameters chosen from region
$A$, the quasi-specific energy measured along the integral accretion solution passing through
the inner sonic point is less than the same measured along the solution passing through the
outer sonic point ($\xi(r_c^{in})<\xi(r_c^{out})$). The situation is just opposite for 
flows characterised
by parameters taken from the region $W$. When parameters are taken from region $W$, the
accretion flow can pass through only one sonic point, however the wind (outgoing) solutions
can have three critical points. Outgoing solutions passing through the inner and the outer
critical points may be joined through a stationary shock. We will, however, not discuss
multi-transonic shocked wind in the present work.

\subsection{General relativistic isothermal shock conditions}

\noindent
We note that $h=1$ for isothermal process, which in turns yield,

\begin{equation}
T^{rr}=\rho\left((v^r)^2+c_s^2 g^{rr}\right)
\end{equation}

\paragraph{NT \& RH discs}

\noindent
In this case the shock-invariant quantity turns out to be

\begin{equation}\label{shockinv_NT_iso}
S_{\rm sh} = \frac{(u^2(1-c_s^2)+c_s^2)}{u\sqrt{1-u^2}}
\end{equation}

\noindent
where we have removed any over all factor of $ r $ as shock invariant quantity is to be evaluated at  $ r=r_{\rm sh} $ for different branches of flow.

\paragraph{ALP discs}

\noindent
In this case, the shock-invariant quantity turns out to be

\begin{equation}\label{shockinv_ALP_iso}
S_{\rm sh} = \frac{(u^2(1-c_s^2)+c_s^2)\sqrt{\lambda^2 v_t^2 - a^2 (v_t -1)}}{u\sqrt{1-u^2}}
\end{equation}

\noindent
where $v_t$ is given in (\ref{v_t}). \\

\begin{figure}[!htbp]
\begin{tabular}{cc}
\includegraphics[scale=0.8]{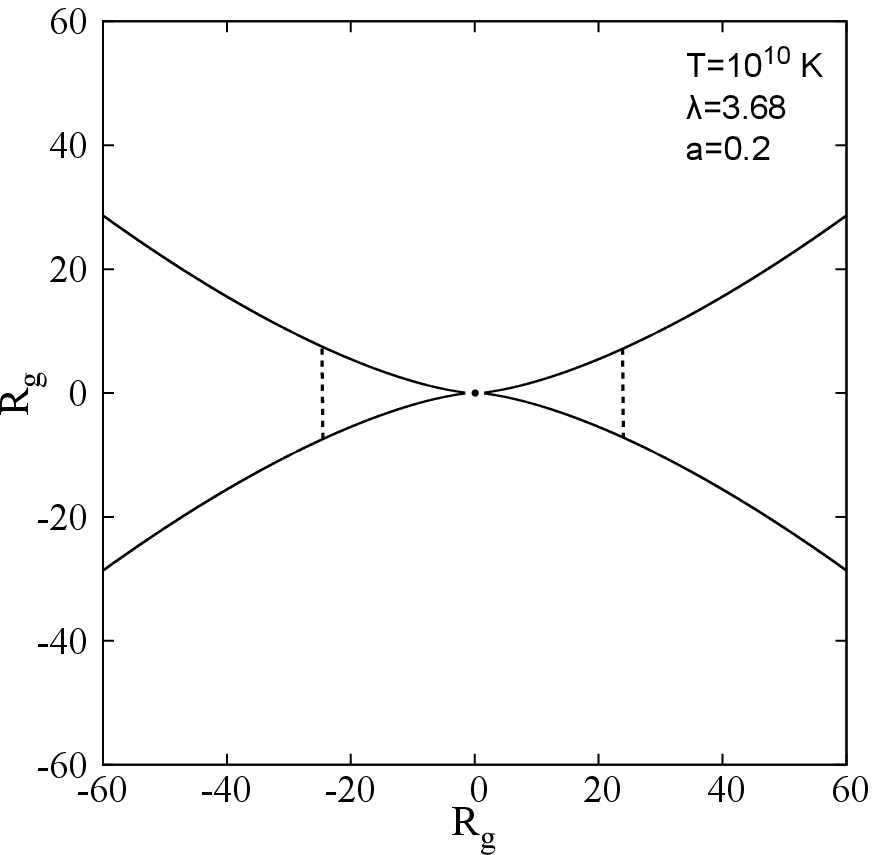} &
\includegraphics[scale=0.8]{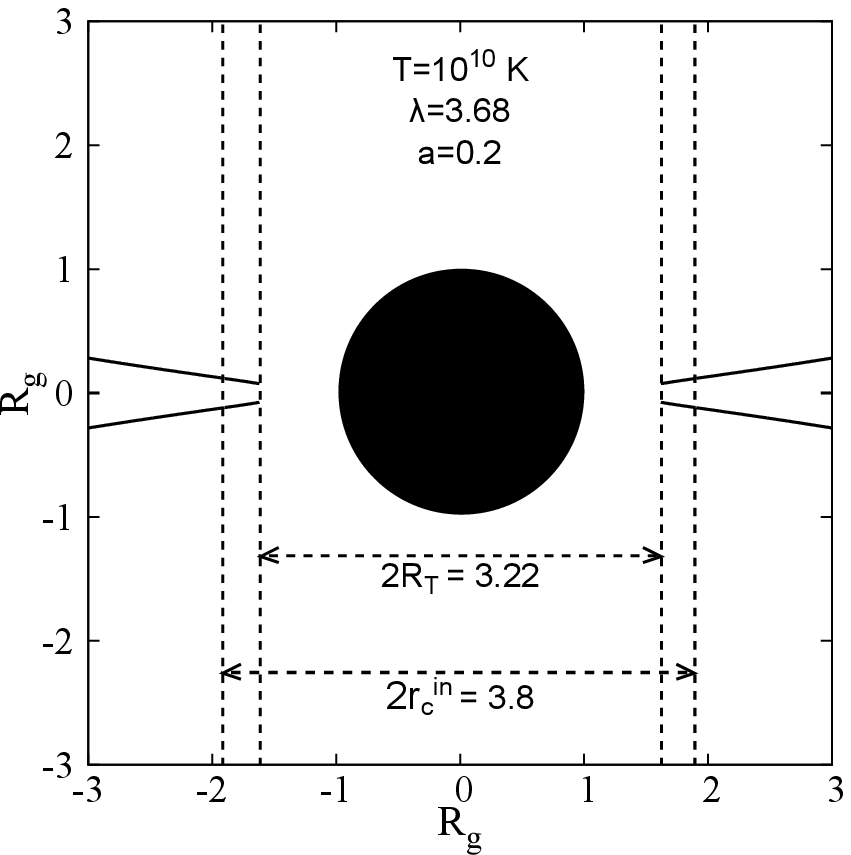} \\
\includegraphics[scale=0.8]{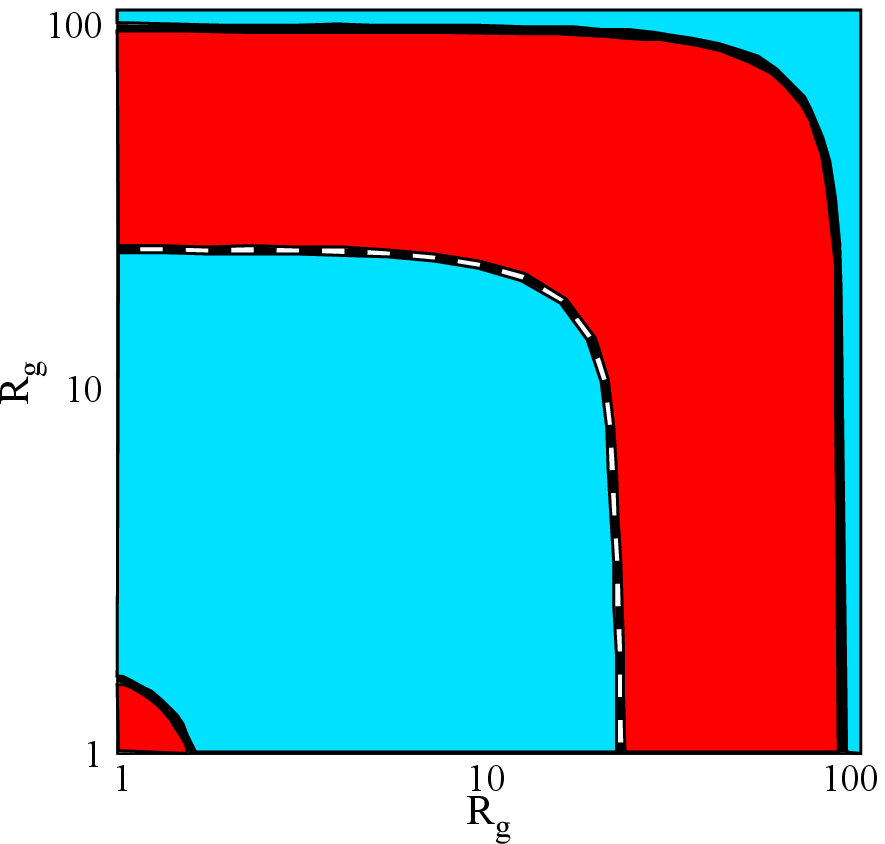} &
\includegraphics[scale=0.8]{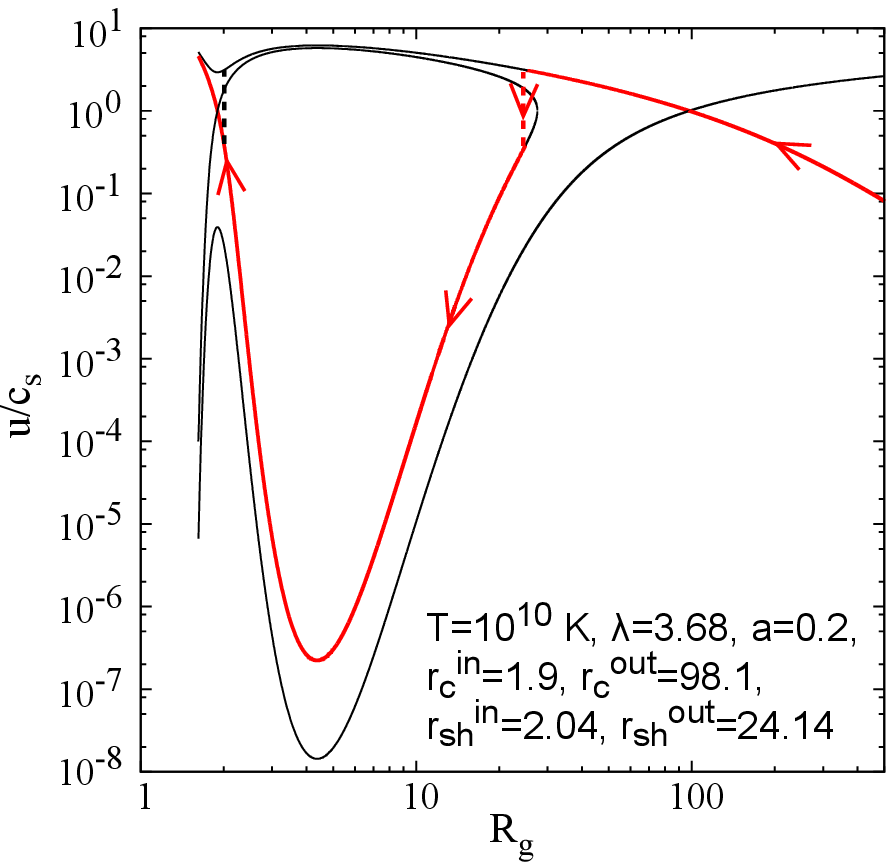} \\
\end{tabular}
\caption{Isothermal NT disc - (a) Disc height vs. radial distance (in units of
$R_g$) with vertical dotted lines depicting the shock locations ($r_{sh} = 24.15$). (b) Magnified
view of the central region depicting the truncation radius ($R_T$) and inner critical point
($r_c^{in}$). (c) Face-on view of the disc (Solid curves represent sonic points and dashed curve
represents shock front. Regions shaded in cyan and red depict subsonic and supersonic flows
respectively). (d) Mach number vs. radial distance profile (red path depicts physical flow in
the direction indicated with arrows).}
\label{fig8}
\end{figure}

\noindent
We show the multi-transonic flow topology with shock for a set of $\left[T,\lambda,a\right]$ 
as specified in the
diagram. We also show the segregated disc structure (the edge-on view in fig.(\ref{fig8}(a)), and the
face-on view in fig.(\ref{fig8}(c))) for various subsonic and supersonic parts of the flow, 
as clarified
in much detail in section $3.4$ while describing features of fig.(\ref{fig2}). There is, however, a major
difference between post-shock disc structure in polytropic flow with energy-preserving shock
and for isothermal flow with temperature-preserving shock. We have seen that for polytropic
shocked accretion, lack of dissipation of energy increases the post-shock flow temperature and
the post-shock part of the disc expands to produce a torus-kind of geometry. For isothermal
shock, however, the thermal energy generated at the shock is allowed to liberate in order
to maintain invariance of the flow temperature. Since no additional thermal energy gets
trapped, the post-shock disc, unlike its polytropic counterpart, does not get puffed-up. The
energy liberated at the shock may power the strong flares emanating out of the axisymmetric
accretion around supermassive black holes at the centre of the galaxies. We shall elaborate
this aspect in subsequent sections. In fig.(\ref{fig8}(b)), the innermost part of the disc has been
shown separately along with the termination radius ($R_T$) of the disc. \\

\begin{figure}[!htbp]
\begin{tabular}{cc}
\includegraphics[scale=0.8]{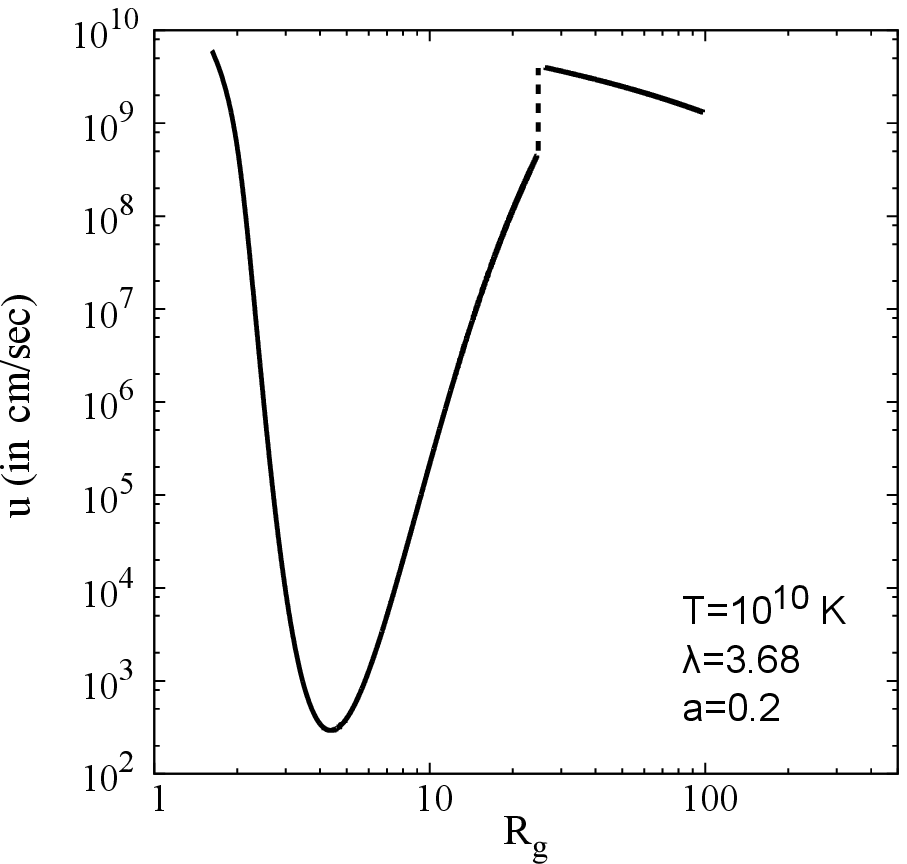} &
\includegraphics[scale=0.8]{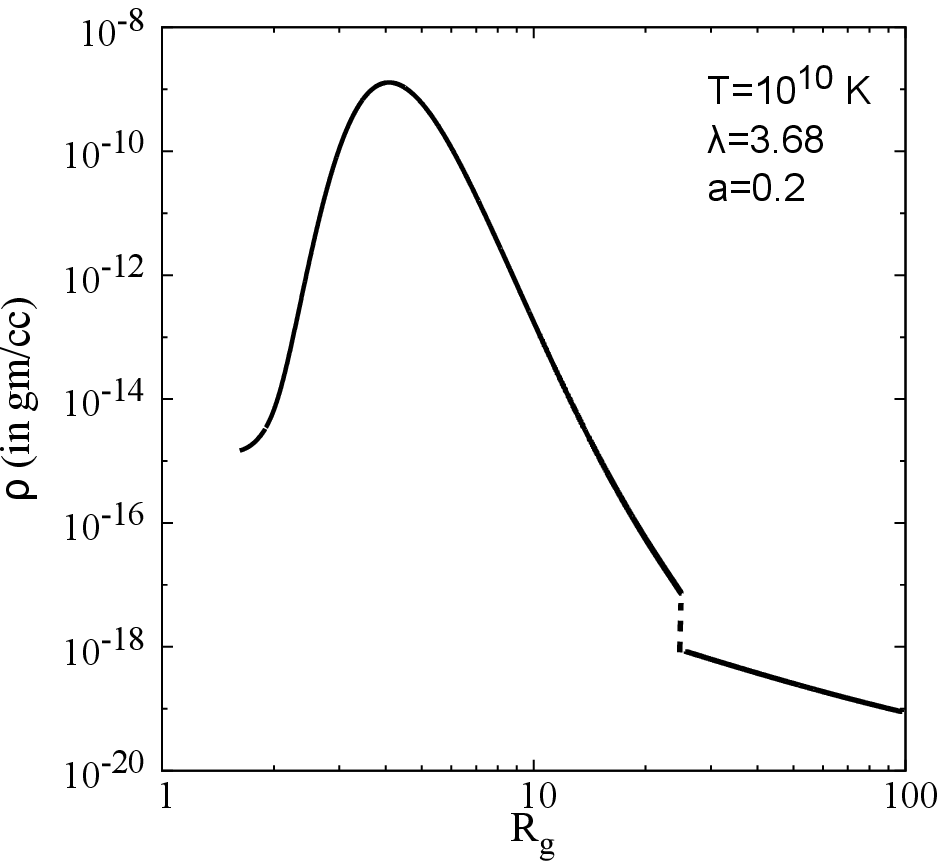} \\
\includegraphics[scale=0.8]{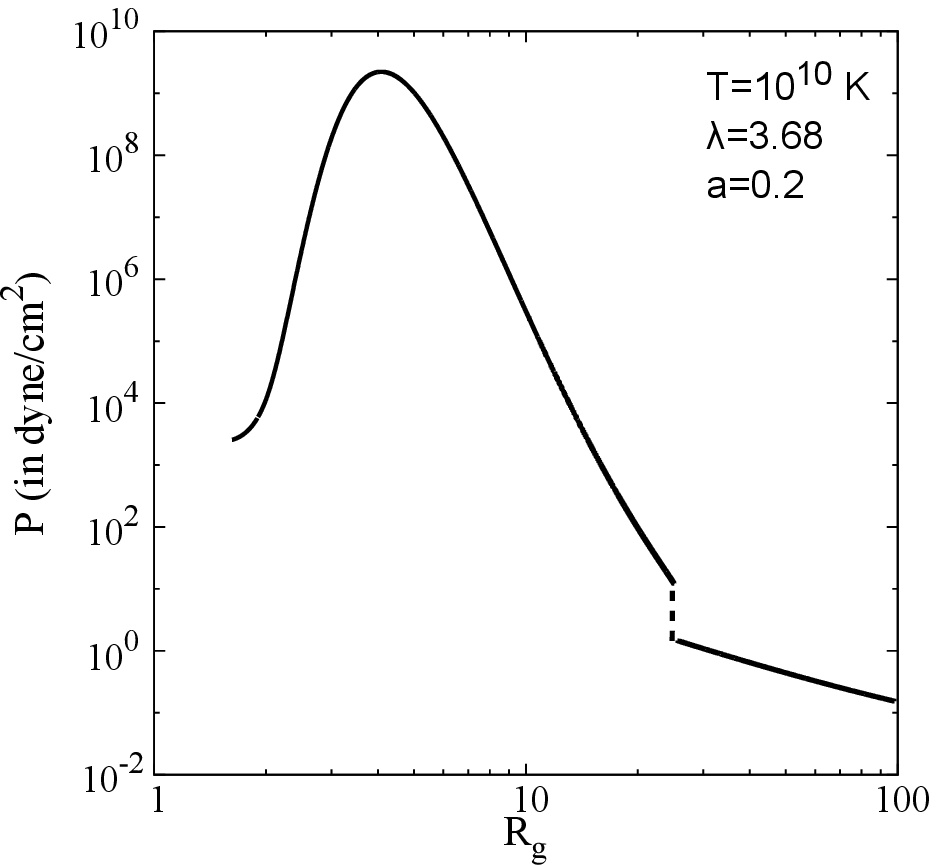} \\
\end{tabular}
\caption{Isothermal NT flow profile - (a) Advective flow velocity ($u$) vs $r$. (b)
Rest mass density ($\rho$) vs. $r$. (c) Pressure ($P$) vs. $r$. $u$ in units of cm/sec, $\rho$ in 
units of gm/cc, $P$ in units of dyn/cm$^2$, $r$ in units of $R_g$.}
\label{fig9}
\end{figure}

\noindent
Fig.(\ref{fig9}) shows the variation of the dynamical velocity $u$, the matter density $\rho$ and 
the fluid
pressure $P$, as a function of the radial distance (measured from the horizon) for the shocked
branch. The vertical dashed line signifies the discontinuous shock transition which joins
the pre-shock flow solution passing through the outer sonic point with the post-shock flow
solutin passing through the inner sonic point. For isothermal accretion, the sound speed
remains invariant, hence the Mach number profile turns out to be just a scaled down version
of the velocity profile. \\

\begin{figure}[!htbp]
\begin{tabular}{cc}
\includegraphics[scale=0.8]{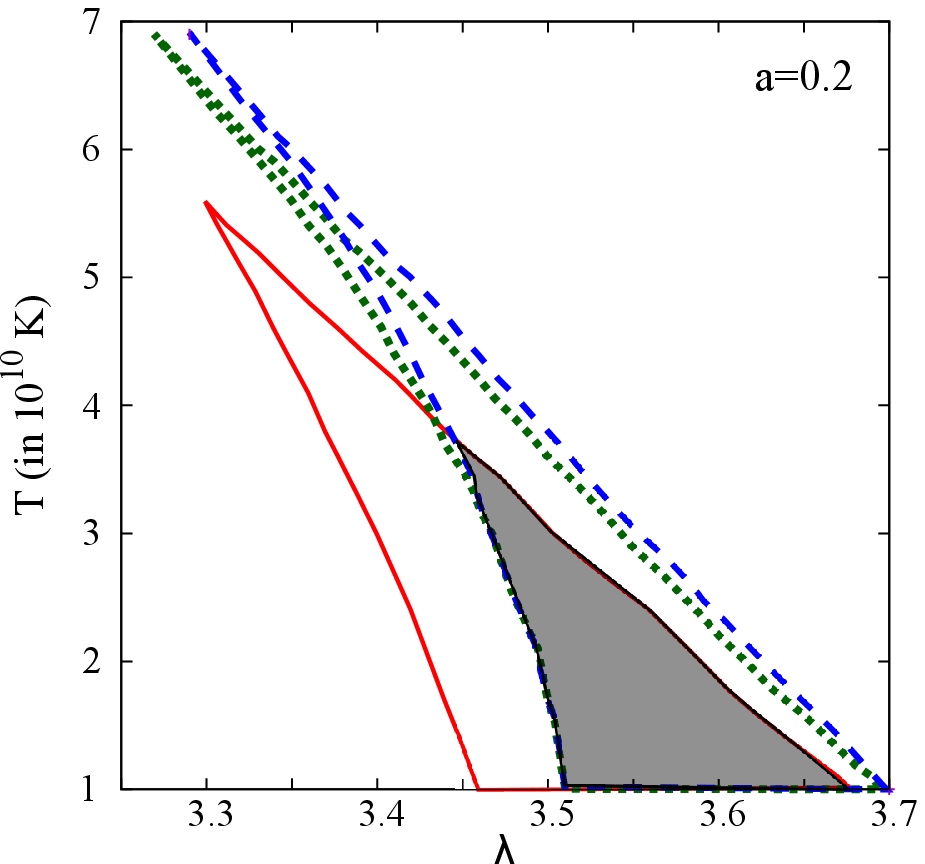} &
\includegraphics[scale=0.8]{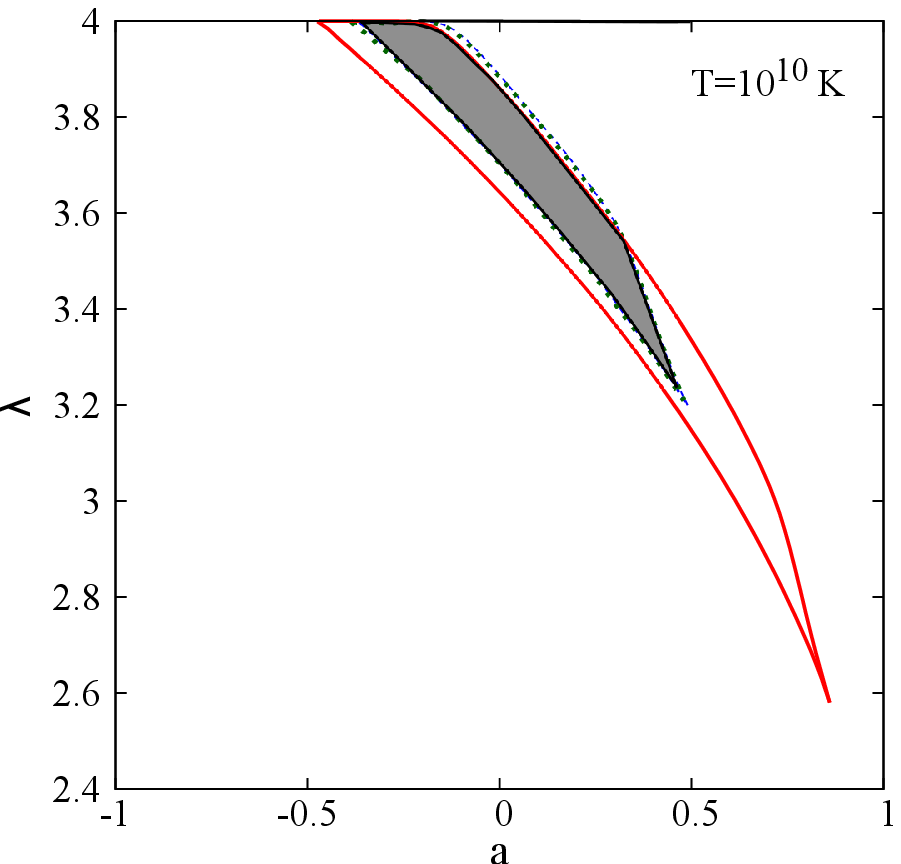} \\
\end{tabular}
\caption{Parameter space for shocked accretion flows - (a) Flow temperature ($T$) vs. $\lambda$
($a = 0.2$). (b) $\lambda$ vs. $a$ ($T = 10^{10} K$). Solid red curves, green dotted curves and 
blue dashed curves represent ALP, RH and NT discs respectively. Shaded regions depict overlap zones
for the three disc models.}
\label{fig10}
\end{figure}

\noindent
Fig.(\ref{fig10}(a)) shows the parameter space (spanned by the flow temperature and flow angular
momentum) for multi-transonic shocked flow for three different disc thicknesses as considered
in our work. The figure has been obtained for a fixed value of black hole spin $a=0.2$. Similar
figures can be obtained for any other value of a for both prograde as well as retrograde flows.
The region of parameter space common to all three disc thicknesses has been shaded in dark grey colour. \\

\noindent
The parameter space spanned by the flow angular momentum $\lambda$ and the black hole spin $a$
has been depicted in fig.(\ref{fig10}(b)), for a fixed value of flow temperature $T=10^{10}$ K. 
This value
is only representative and similar diagrams with the same general features can be obtained
for other values of $T$ as well. The particular value of $T$ has been chosen to cover an extended
range of $a$ allowing shock solutions. Similar to the polytropic case (fig.(\ref{fig4}(b))), 
lower values of the Kerr parameter permit shock formation for flows with higher values of the 
specific angular momentum. \\

\begin{figure}[!htbp]
\begin{tabular}{cc}
\includegraphics[scale=0.8]{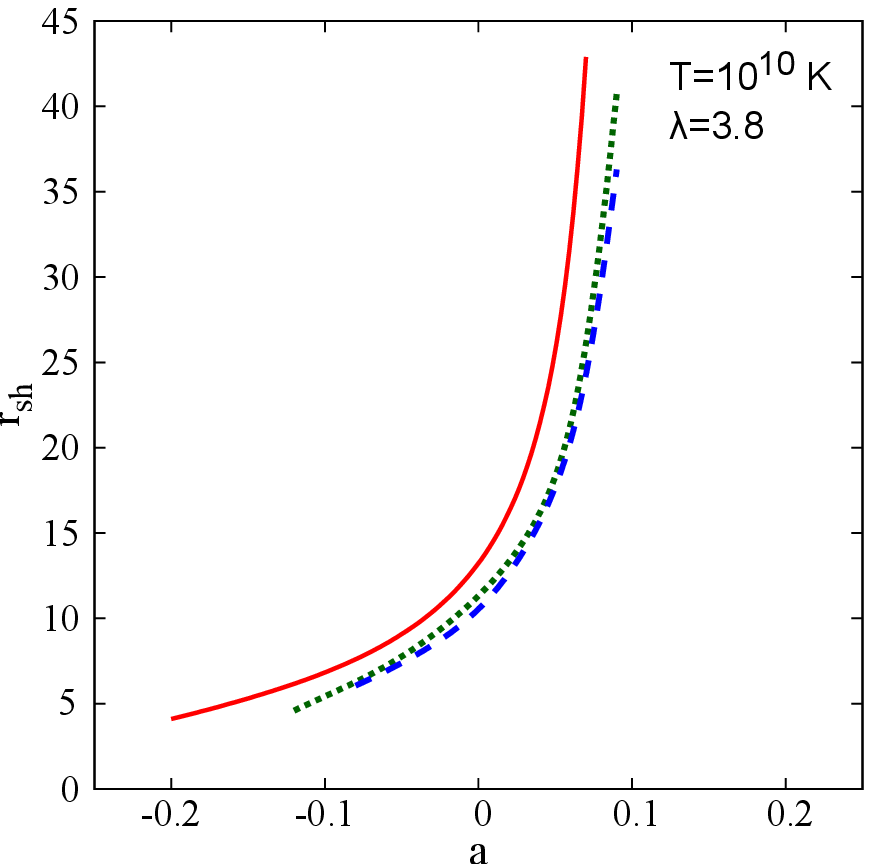} &
\includegraphics[scale=0.8]{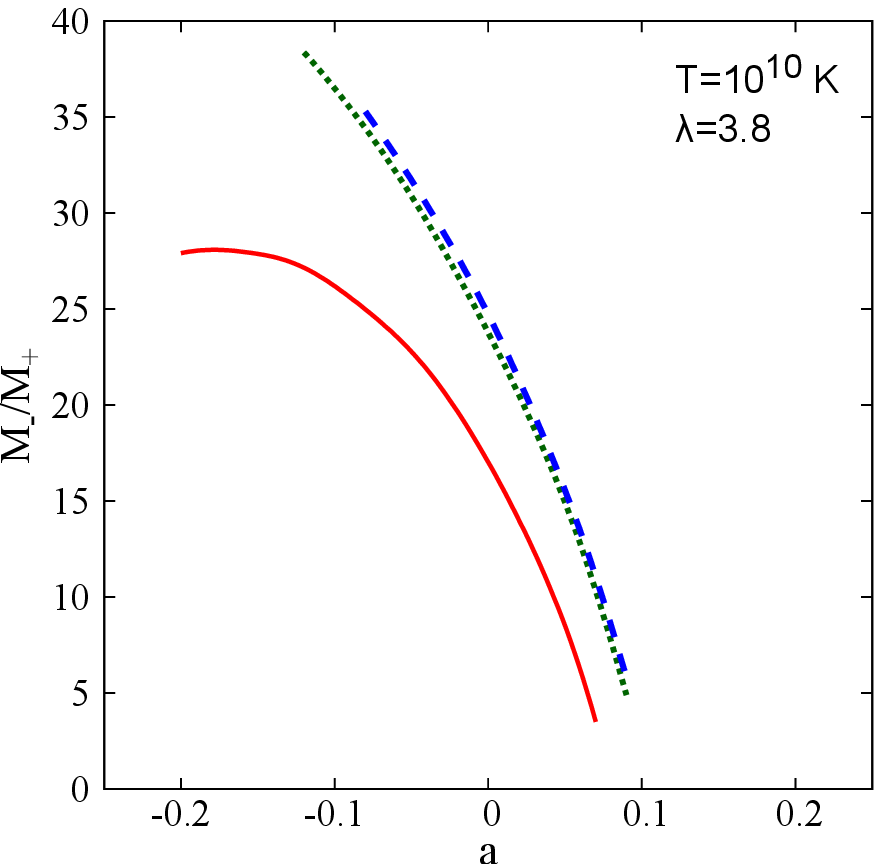} \\
\includegraphics[scale=0.8]{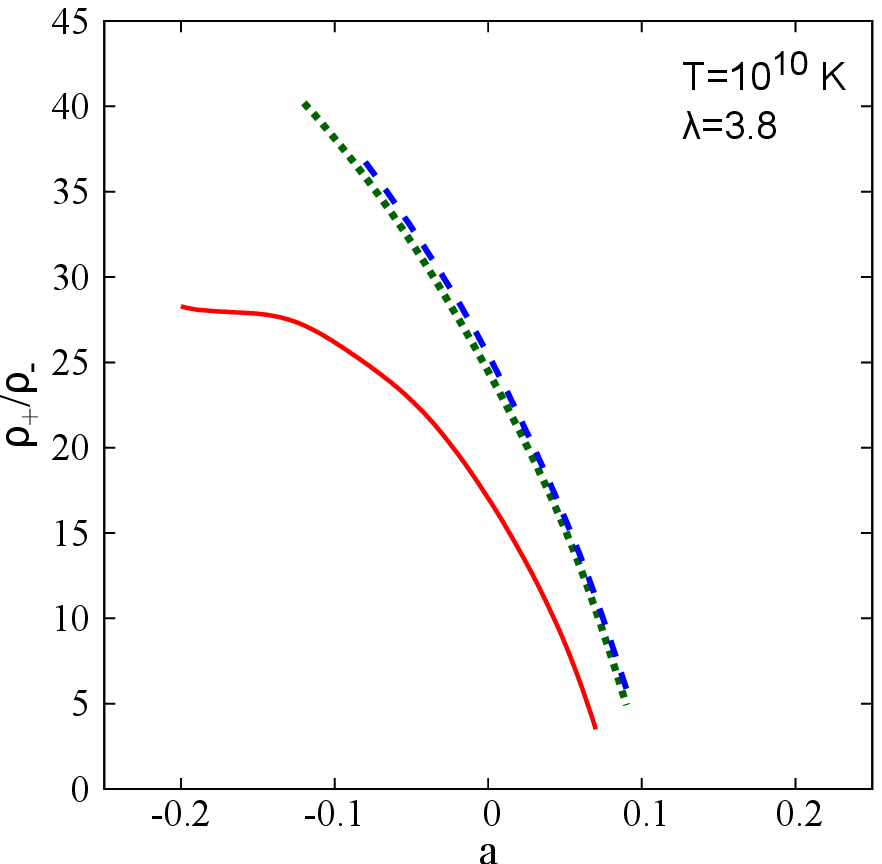} &
\includegraphics[scale=0.8]{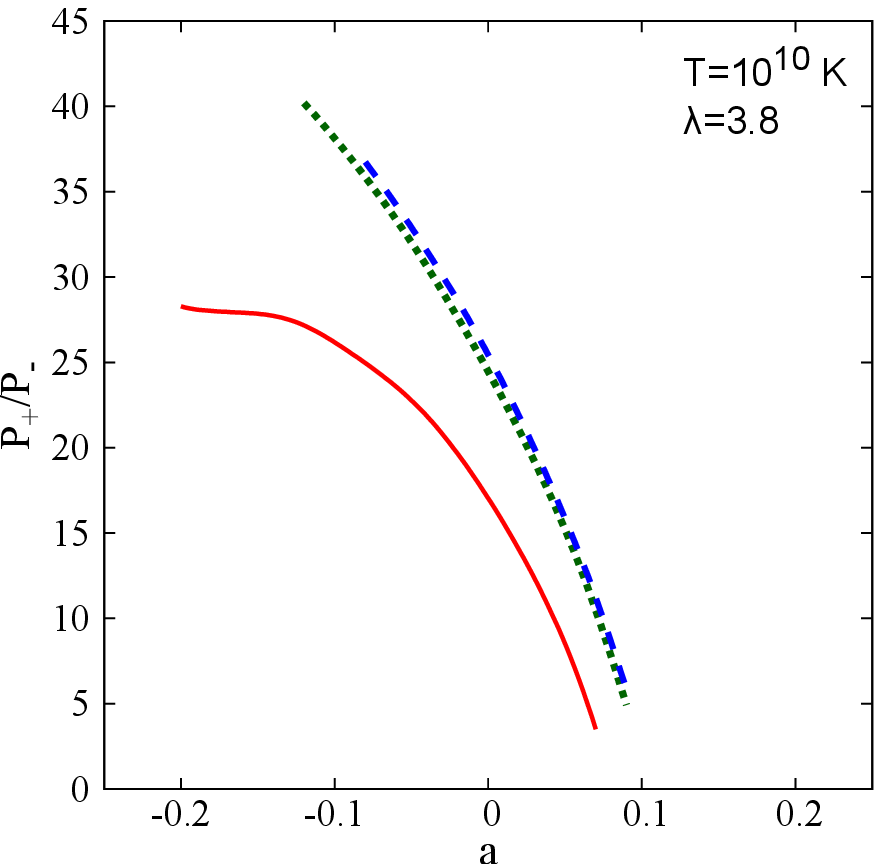} \\
\end{tabular}
\caption{Shock location - (a) $r_{sh}$ (in terms of $R_g$) vs. $a$, Ratios at shock - (b) $M_-/M_+$
vs. $a$, (c) $\rho_+/\rho_-$ vs. $a$ and (d) $P_+/P_-$ vs. $a$. `-' and `+' refer to `before' and `after' shock
respectively. ALP shown by red solid lines, RH shown by green dotted lines and NT shown
by blue dashed lines.}
\label{fig11}
\end{figure}

\noindent
Fig.(\ref{fig11}(a)) shows the variation of shock location with the black hole spin 
for both the co-rotating
as well as counter-rotating flows. Faster rotating black holes produce the shock at
larger distances for prograde flow whereas the trend is reverse in case of retrograde accretion.
It should be noted that the NT and RH-type discs produce shocks in extremely nearby locations
for isothermal flows. Similar conclusions drawn while observing the variation of the
ratios of the pre-(post-) to the post-(pre-) schock accretion variables as a function of the Kerr
parameter. Such variations are shown in fig.(\ref{fig11}(b-d)). The shock becomes stronger and the
post-shock flow becomes denser, as the shock location approaches towards the horizon. This
is physically consistent as larger amounts of gravitational potential energy will be available
for liberation when shock forms closer to the horizon. The results have been obtained
for a particular set of $\left[T,\lambda\right]$. But results with the same variational 
trends can be obtained
for any other relevant values of the given flow parameters. 

\subsection{Energy dissipation at temperature-preserving shock}

\noindent
The quasi-specific energy $\xi$ plays a role similar to that played by entropy accretion rate $\dot{\Xi}$
for polytropic flow. $\xi$ decreases after the flow encounters a shock. The difference of values of
$\xi$ computed along the integral accretion solutions passing through the outer and the inner
sonic points, respectively, is a measure of the flow energy liberated at shock. Such energy
liberation mechanism may explain the formation and dynamics of flares (as observed in various
wavelengths) emanating out from the proximity of our own Galactic Centre black hole. \\

\begin{figure}[!htbp]
\begin{tabular}{cc}
\includegraphics[scale=0.7]{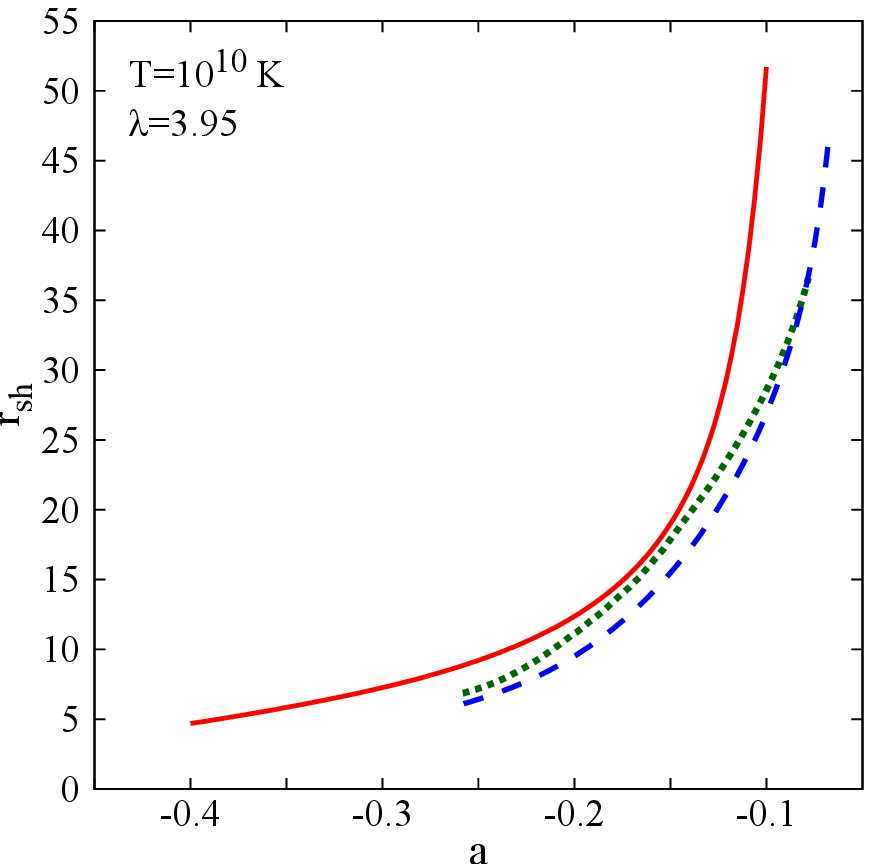} &
\includegraphics[scale=0.7]{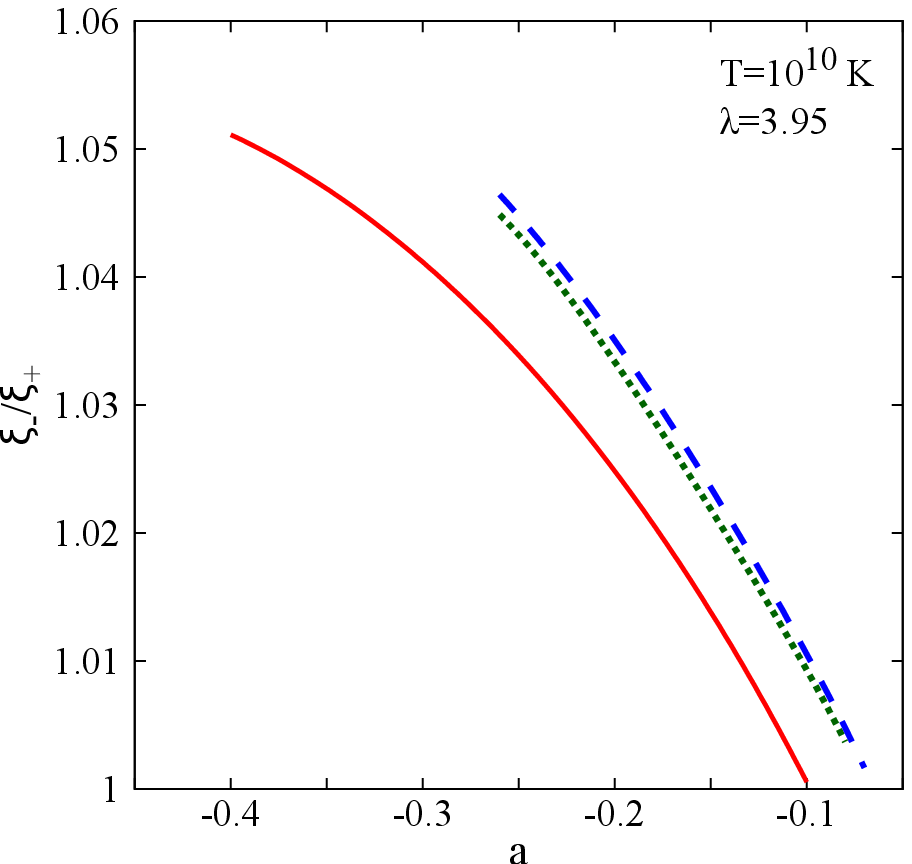} \\
\includegraphics[scale=0.7]{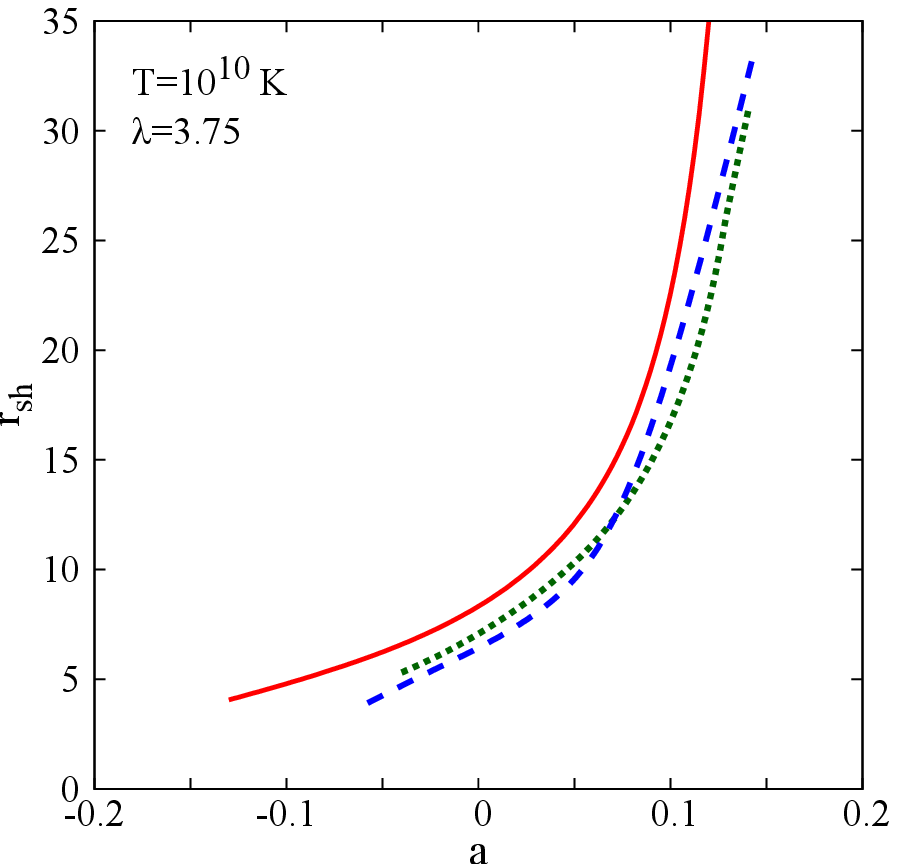} &
\includegraphics[scale=0.7]{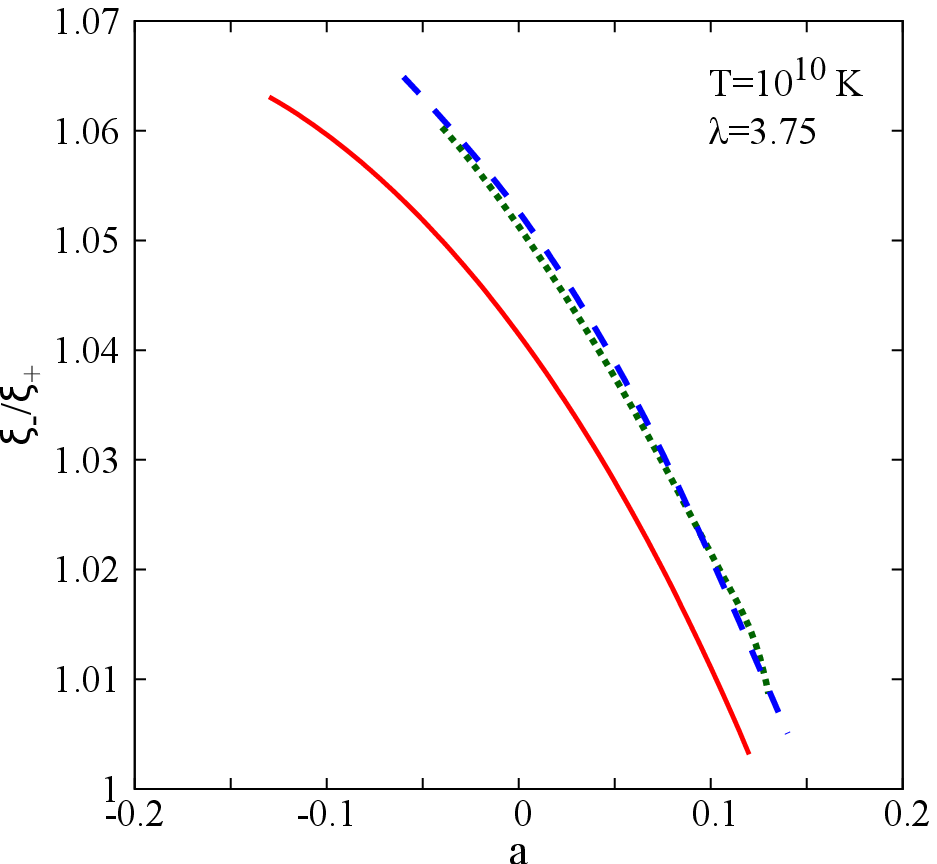} \\
\includegraphics[scale=0.7]{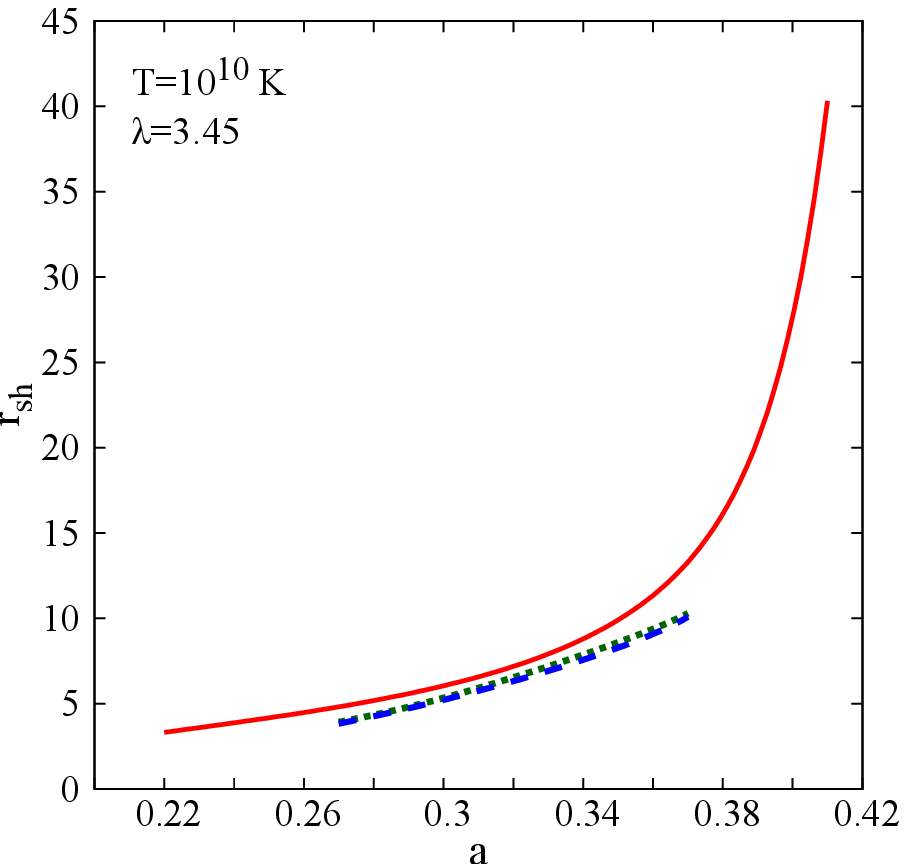} &
\includegraphics[scale=0.7]{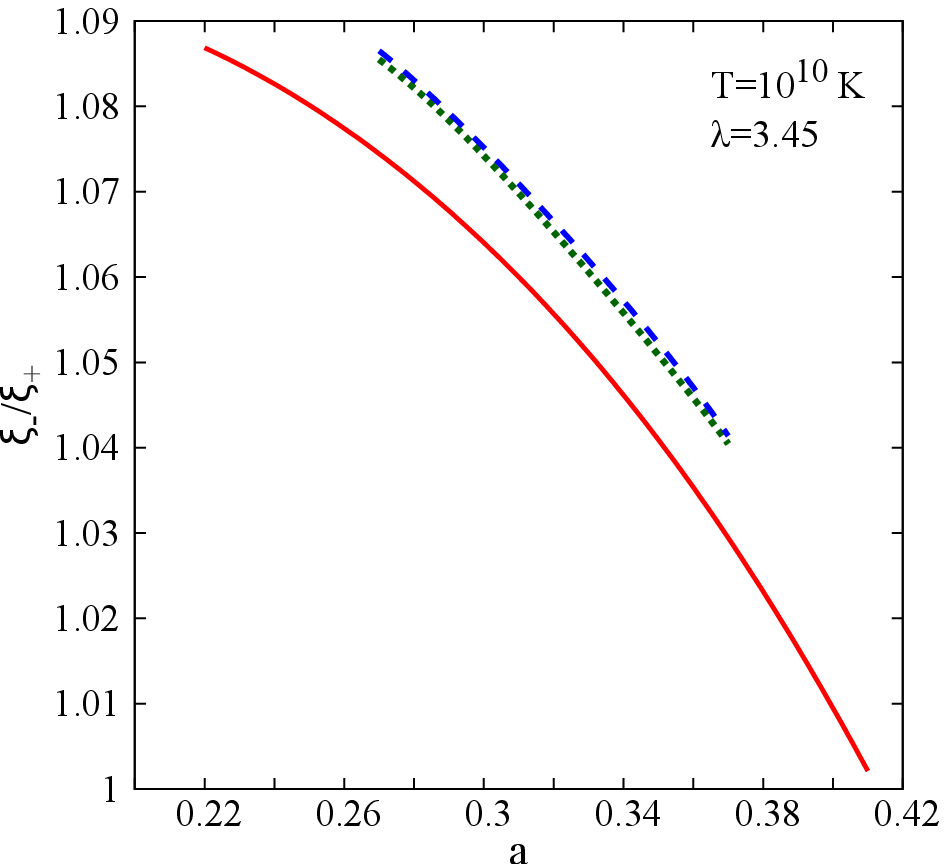} \\
\end{tabular}
\caption{Energy dissipation at shock - $\xi_-/\xi_+$ vs. $a$ along with the corresponding $r_{sh}$ (in
terms of $R_g$) vs. $a$ for $T = 10^{10}$K and (a) $\lambda = 3.95$, (b) $\lambda = 3.75$ and (c) $\lambda = 3.45$. `-' and
`+' represent quantities `before' and `after' shock respectively. ALP shown by red solid lines,
RH shown by green dotted lines and NT shown by blue dashed lines.}
\label{fig12}
\end{figure}

\noindent
In fig.(\ref{fig12}), we plot the ratio of the pre- to post- shock values of $\xi$ as a function of 
the Kerr parameter. As a reference, we also show the variation of corresponding shock locations with
the black holes spin. Three sets of figures have been produced for three different values of
the flow angular momentum $\lambda$ and for the same value of the temperature as shown in the
figure. \\

\noindent
We expect that the ratio of $\xi$ at shock might anti-correlate with the shock location $r_{sh}$,
since for smaller values of $r_{sh}$ (forming closer to the horizon, in a relatively stronger gravity
regime), the gravitational potential energy available for liberation is higher. Also, values of
the effective centrifugal barrier, i.e. ($\lambda \pm a$) determine the amount of energy dissipated at
the shock. For lower values of ($\lambda \pm a$), accretion flow has larger values of 
the radial advective
velocity. This velocity is directed, and gets randomised at the shock. The larger the value of
the directed bulk velocity, the higher is the amount of energy liberated when it gets randomised 
through shock formation. The value of the ratio of the pre- to post-shock quasi-specific
energy, should, thus anti-correlate with $\lambda$, as well as with ($\lambda \pm a$). \\

\noindent
This is exactly what we observe in the figure (see three consecutive panels a--c in fig.(\ref{fig12})).
We also see that the amount of energy liberated can be as large as (approximately) 9\%.
Thus the disc may become considerably luminous (on the corresponding wavelength) at the
shock, and can also produce a radiatively effcient post-shock flow. It has been found (see
the figures) that among the three different disc-height recipes, the NT-type of disc liberates
maximum amount of energy, and hence becomes maximally luminous at the shock, provided
the initial set of boundary conditions remain the same. \\

\noindent
It also requires to be mentioned that the aforementioned energy-liberation process is not
similar to the Blandford-Znajek (BZ) mechanism (\citep{bz77mnras}), where the rotational energy
of the black hole is extracted to power jets. BZ mechanism requires the presence of poloidal
magnetic field lines around a spinning black hole, which extracts the rotational energy of
the hole itself. On the contrary, our simple theoretical model of purely general relativistic
hydrodynamic flow does not include any magnetic energy component. The energy-liberation
mechanism discussed in our work is not similar to the Penrose process (\citep{pf71nps}) as well,
since in our model, energy gets liberated at the shock location residing well outside the
ergosphere.

\section*{Acknowledgement}
PT would like to acknowledge the kind hospitality provided by
Harish-Chandra Research Institute, HBNI,
Department of Atomic Energy, Government of India, for
supporting his visit through the plan budget of Cosmology
and High Energy Astrophysics grant.

\bibliographystyle{plainnat}
\bibliography{paper1}

\end{document}